\documentstyle[preprint,eqsecnum,aps]{revtex}
\tighten
\begin{document}
\preprint{}
\draft
\def\ab{$(n+m)$}
\def\cd{$(n+1)$}
\title {Split structures in general relativity and \\
        the Kaluza-Klein theories}
\author{V.D. Gladush\footnote{E-mail: gladush@ff.dsu.dp.ua},
            R.A. Konoplya }
\address{Department of Physics, Dnepropetrovsk State University, \\
per. Nauchny 13, Dnepropetrovsk, 320625 Ukraine}
\date{\today}
\maketitle

\begin{abstract}

We construct a general approach to decomposition of the tangent bundle
of pseudo-Riemannian manifolds into direct sums of subbundles, and the
associated decomposition of geometric objects. An invariant structure
${\cal H}^r$ defined as a set of $r$ projection operators is used to induce
decomposition of the geometric objects into those of the
corresponding subbundles. We define the main geometric objects characterizing
decomposition. Invariant non-holonomic generalizations of the
Gauss-Codazzi-Ricci's relations have been obtained. All the known types
of decomposition (used in the theory of frames of reference, in the
Hamiltonian formulation for gravity, in the Cauchy problem, in the theory
of stationary spaces, and so on) follow from the present work as special
cases when fixing a basis and dimensions of subbundles, and
parameterization of a basis of decomposition. Various methods
of decomposition have been applied here for the Unified Multidimensional
Kaluza-Klein Theory and for relativistic configurations of a perfect fluid.
Discussing an invariant form of the equations of motion we have found
the invariant equilibrium conditions and their 3+1 decomposed form.
The  formulation of the conservation law for the curl has been obtained
in the invariant form.
\end{abstract}
\narrowtext

\section{INTRODUCTION}

Most approaches and formalisms in General Relativity as well as in the
multidimensional Unified Theories are connected with decomposition of
spaces into direct sums of subspaces and the associated decomposition
of geometrical objects. It means that in addition to usual
structures (differentiable structure, the metric structure, and so on) one
should introduce {\bf a split structure} which induces the decomposition of
manifolds. This extra structure determines decomposition of all objects and
structures defined on a manifold. Among varieties of formalism of
decomposition are the methods
aimed to describe frames of reference and observable quantities in the theory
of gravity. Similar methods have gained the wide acceptance in a great number
of problems. Some of these problems are the canonical formalism for
gravitational waves, the Cauchy problem in General Relativity, construction
of the Unified Theory of interacting fields, quantization of the gravitational
field, the tetradic formalism, the Newman-Penrose formalism, the theory of
stationary and axisymmetric gravitational fields, the multidimensional and
four-dimensional cosmologies.

Mathematicians and physicists developed methods of decomposition starting
mainly from
their intrinsic interests. It often took place independently and parallely,
so that sometimes the same advances were overlooked and then refound.

The early stage of the development of mathematical technique for decomposition
could be seen in the classical theory
of hypersurfaces, and in the theory of $n$-dimensional surfaces imbedded
in the $n$-dimensional Riemannian manifold ~\cite{Eisa}. Then, in the history
of split methods, we can distinguish several ways.
In mathematics, at the classical stage,
there were constructed coordinate techniques for non-holonomic spaces and
subspaces ~\cite{Schout}, \cite{Vagn}. Owing to the use of the
coordinate language such methods were rather cumbersome. In physics split
methods were induced by attempts to create the Unified Theory of fields, and,
in particular, by appearance of the Kaluza-Klein theory ~\cite{Kaluz},
\cite{Klein}. This led to construction of
a $4+1$ split method for a $5$-dimensional manifold and gave an impulse
to study multidimensional Kaluza-Klein theories and multidimensional
cosmologies ~\cite{DeSab}-\cite{Ivas} (see also references in~\cite{Ivas}).

Another physical branch of split methods was developed much more later than
the original works on a $4+1$ split were. This branch has begun, apparently,
 with the work by Echart ~\cite{Echart} and has been completed in the papers
~\cite{Zel1}-\cite{Pol}.
There have been constructed $(3+1), (2+2), (1+1+2)$ coordinate split methods
and their special cases (\cite{Ant} and references therein).

The other independent direction to construct a split of spaces in
General Relativity is connected with the questions of convenience of
mathematical representation of the Einstein equations and with the study of
these equations' structure. This branch is brought about by needs for
construction of the canonical formalism ~\cite{ADM}, of various projection
formalisms in the theory of the stationary and axisymmetric gravitational
fields ~\cite{Ger}-\cite{Kram} and for the positing of the Cauchy problem in
General Relativity ~\cite{Lich}.

Unfortunately, many of the works mentioned above, which have already become
classical, use different and often inassociated approaches.
Moreover a coordinate language applied there, especially in early works,
makes it almost impossible to calculate the Einstein equations for some forms
of the metric.

New stage of development of split methods is based on modern differential
geometry ~\cite{Ster}, \cite{Kob}. Its invariant language have become working
one in General Relativity ~\cite{Haw}. It is not only a natural language
for geometry
in the whole, but also a convenient approach to calculations. Obtaining
formulae is reduced, the formulae themselves become universal, and all the
calculations can easily be made by computer.

The invariant split method was considered in ~\cite{Gray} but without any
connection with the previous works on a split. Objects introduced formally in
this work
have no clear geometrical meaning. One of us proposed the general invariant
method of an $(n+m)$ split for pseudo-Riemannian manifolds
~\cite{Gl1}-\cite{Gl4}. There were most approaches to split
unified in the works, and the objects introduced there have clear physical
and geometrical meaning. For special cases of
$(1+4),~(1+3),~(2+2),~(n+4)$ splits, in the coordinate representation,
these objects reduce to known physical characteristics of
a system ~\cite{Vlad1}-\cite{Pol}.

Multidimensional cosmologies and the Unified Theories imply that a manifold
should be decomposed into more than two submanifolds. That is one reason
why a split of a manifold requires the most general representation.

The theory of $(n_1+n_2+...+n_r)$ decomposition of a pseudo-Riemannian
manifold into the $r$ non-holonomic orthogonal subbundles
$\Sigma_a$ of a dimension $n_a~~(a=1,2,...r)$ has been constructed in the
present work ($n=n_1+n_2+...+n_r$ is dimension of a manifold).
The $(n+m)$ and $(n+1)$ forms of invariant decomposition have obtained as
consequences. Choosing the projection operators and gauges of a basis of
decomposition we construct various special cases. Some applications of this
method are considered. Let us emphasize that we don't refer to problems of
global geometry, but use its invariant formulations to construct
decomposition of spaces into direct sums of subspaces.

Note, that the theory of structures mentioned above has found its further
mathematical development, and now is widely known in differential geometry
under the name "almost product structure" \cite{Gray}. The latter can be
treated from " G-structure" point of view \cite{Cher} (see also \cite{Bor}
and references therein).  We will follow a more natural approach and
use the term "a split structure", which is, in our opinion, more in
the spirit of physical conceptions aroused
in General Relativity when dealing with $(3+1), (2+2), (1+1+2)$
decomposition of space-time.

The plan of the paper is as follows. In SecII we introduce the necessary
notations used in differential geometry, and the main definition of the
theory, {\bf a split structure} on a pseudo-Riemannian manifold. Then we
introduce
the metrics and connections induced  on  the  corresponding  subbundles  as
well as the main associated geometrical  objects  on  the  subbundles:
the  tensors of  extrinsic curvature and  of extrinsic torsion, analogies
of the Ricci coefficients  of  rotation, and  the  curvature tensor.
The invariant non-holonomic
generalization of the Gauss-Codazzi-Ricci's relations has been found as
various projections of the curvature tensor into every possible subbundle
(see Appendix A).

In SecIII the special case of invariant split structure $r=2$ (when we deal
with two subbundles only) is considered. In this case the generalized
coefficients of rotation disappear from the curvature tensor, thereby the
final formulae become much simpler.
In SecIV the invariant formulae in the $(n+1)$ split form complete the scheme of
the invariant split for a pseudo-Riemannian manifold. Further,
for any concrete calculations, we must fix projection operators.

In SecV we, for illustration, briefly consider the $(n+m)$ and $(n+1)$
coordinate decomposition of the manifold with respect to the natural basis
$\{\partial_{\mu}, dx^\nu\}$. In SecVI the method of $(n+m)$ decomposition is
constructed with respect to an adopted basis. All the relations
obtained in SecVI are basic ones for the other variants of decomposition in
this paper. The final formulae (in Appendixes B,C) can be used as an algorithm
to compute the Ricci tensor, the Riemann tensor, the scalar curvature, and
the corresponding Lagrangians.

In SecVII we define the canonical parameterization of a basis of decomposition. There
have been obtained the main geometric objects with respect to this basis.
Various well known
special cases, which follow from this para\-metrization, are discussed in the
section. Connections and relations among them are analyzed.

In SecVIII we obtain the decomposition induced by a given family of surfaces.
In SecIX
we consider the decomposition induced by a group of motion. On the basis of
this section's results we construct Lagrangian of the Unified multidimensional
Kaluza-Klein Theory (SecX). This decomposition, apart from everything else, serves as a
methodical illustration of the possible use of the present method for physical
theories.

Finally SecXI deals with the theory of configurations of a perfect fluid. Using the
$(3+1)$ canonical parameterization, one can define a one-form of the enthalpy
and a two-form of the curl. We have obtained the invariant equations of
motion for a perfect fluid. The conservation law for the curl of an isentropic
perfect fluid has been obtained in the invariant form. For this fluid,
rotating in the stationary gravitational field, we have also deduced the
equilibrium conditions and constructed its Lagrangian.

In this work we considered a torsion-free pseudo-Riemannian manifold only, none
the less, our approach can be used without principal changes for theories of
gravity with non-zero torsion. We see further development of the present theme
in the possible expanding of the invariant decomposition to supergravity
theories. We, mostly, used notations and definitions of the works
\cite{Ster}~-\cite{Haw}.

\section{A Split Structure on a pseudo-Riemannian manifold}

Let $M$ be a pseudo-Riemannian manifold with the metric $g$;
$T(M) = {\bigcup_{p\in M}}T_p$ and $T^{*}(M) = {\bigcup_{p\in M}} T^{*}_p$
are the tangent and cotangent bundles over $M$, where $T_{p}$ and $T^{*}_{p}$
are the corresponding fibres over a point $p$ of $M$. The objects
$X,Y,Z,...\in T(M)$ and $\alpha,\beta,\omega,df\in T^{*}(M)$ denote
contravariant and covariant vector fields ($d$ is an exterior differential).
We shall denote by $\omega(X)$ an inner product of a one-form $\omega$
and vector $X$. The scalar product of two vectors $X,Y,$ and two forms
$\alpha,\beta$ is determined by the metric $g$
\begin{equation}\label{1}
X\cdot Y\equiv (X,Y)\equiv g(X,Y);~~~~~~
<\alpha,\beta>\equiv g^{-1}(\alpha,\beta)
\end{equation}
where $g^{-1}$ is the inverse of the metric $g$.

We need to note that for each vector field $Y\in T(M)$ a dual one-form $\omega$
is determined uniquely by $\omega(X)=g(X,Y),~\forall X\in T(M)$.
From now on we just will write $\omega=g(~.~,Y)$. Then the
inverse of the metric $g$ is given by
\begin{equation}
   g^{-1}(\omega,\alpha)=g^{-1}(g(~.~,Y),\alpha)=\alpha(Y),~~~~~~
\forall Y\in T(M),~~~\forall \alpha\in T^{*}(M)
\end{equation}
so that $Y=g^{-1}(~.~,w)$.

A linear operator $L$ on $T(M)$ is a tensor of type $(1,1)$ which acts
according to the relation $L\cdot X\equiv L(X)\in T(M),~~\forall X\in T(M)$.
Then
\begin{equation}
(L^{T}\cdot\omega)(X)=(\omega\cdot L)(X)\equiv\omega(L(X)),
~~~~~\forall X\in T(M)
\end{equation}
where $L^{T}$ is a transpose of an operator $L$.

The product of two linear operators $L\cdot H$ is defined by
\begin{equation}\label{2}
(L\cdot H)\cdot X=L\cdot(H\cdot X)\in T(M), ~~~~~~\forall X\in T(M).
\end{equation}

An operator $H$ is called a symmetric one if
\begin{equation}\label{3}
(H\cdot X,Y)=(X,H\cdot Y), ~~~~~~\forall X,Y\in T(M).
\end{equation}

We have to introduce the new notation, {\bf a split}, which denotes
decomposition into direct sums.
Therefore we shall say that {\bf a split structure} ${\cal H}^r$ is
introduced on $M$ if the $r$ linear symmetric operators $H^a (a=1,2,...r)$
of a constant rank with the properties
\begin{equation}\label{4}
H^a\cdot H^b=\delta^{ab}H^b; ~~~~~~\sum_{a=1}^{r}H^a=I,
\end{equation}
where $I$ is the unit operator $(I\cdot X=I,~~\forall X\in T(M))$, are
defined on $T(M)$.

Now we introduce the notations:
\begin{equation}\label{5}
\Sigma_{p}^{a}\equiv {\rm Im}~H^{a}_{p};~~
(\Sigma_{a}^{*})_{p}\equiv {\rm Im}~(H^{a}_{p})^T;~~
n_a={\rm dim} \Sigma_{p}^{a} = {\rm dim} (\Sigma_{a}^{*})_{p}
\end{equation}
where ${\rm Im}~H^{a}_{p}$ is an image of an operator $H^a$ at a point $p$ of $M$,
i.e. $\Sigma_{p}^{a}=\{X_p\in T_p\mid H^a\cdot X_p=X_p\}$.
It is important that owing to constancy of a rank of the operator $H^a$,
dimension $n_a$ does not depend on a point $p$ of $M$.

From the definitions presented here we can obtain the decomposition of the
tangent and cotangent spaces:
\begin{equation}\label{6}
T_p = {\bigoplus_{a=1}^{r}}\Sigma_{p}^{a};~~~
T^{*}_p = {\bigoplus_{a=1}^{r}}(\Sigma_{a}^{*})_p;~~~~
{\rm dim}~T_p = {\rm dim}~T^{*}_{p} = \sum_{a=1}^{r}n_a
\end{equation}
where the sign $\oplus$ denotes the direct sum. Thus the tensors $\{H^a\}$
are the projection operators, which bring about  decomposition of the fibres
$T_p,~T^{*}_{p}$ into the $r$ local subspaces $\Sigma_{p}^{a}$ and
$(\Sigma_{a}^{*})_p$ respectively. By the same way, the bundles $T(M)$ and
$T^{*}(M)$ are decomposed into the $(n_1+n_2+...+n_r)$ subbundles
$\Sigma^{a}$, $\Sigma_{a}^{*}$, so that
\begin{eqnarray} \label{7}
    T(M) = {\bigoplus_{a=1}^{r}}\Sigma^{a};~~~~
T^{*}(M) = {\bigoplus_{a=1}^{r}}\Sigma^{*}_{a};~~~~
\Sigma^a = {\bigcup_{p\in M}}\Sigma_{p}^{a};~~~~
\Sigma^{*}_a = {\bigcup_{p\in M}}(\Sigma_{a}^{*})_p.
\end{eqnarray}

Then arbitrary vectors, covectors, and metrics are decomposed according
to the scheme:
\begin{equation}\label{8}
X = \sum^{r}_{a=1} X^a,~~~
\alpha = \sum^{r}_{a=1} \alpha_a,~~~
g = \sum^{r}_{a=1} g^a,~~~
g^{-1} = \sum^{r}_{a=1} g^{-1}_a
\end{equation}
where
\begin{equation}\label{9}
X^a = H^a\cdot X^a = H^a\cdot X;~~~
H^b\cdot X^a = 0;~~~
X^a\cdot X^b = 0;~~~(a\not= b)
\end{equation}
\begin{eqnarray}\label{10}
     & \alpha_a = \alpha_a\cdot H^a = \alpha\cdot H^a;~~~
\alpha_a\cdot H^b = 0;~~~
\alpha_a (X^b) = 0~~~(a\not= b)  \nonumber\\
     & g^{a}(X^a,Y^a)\equiv g(X^a,Y^a);~~~
g^{-1}_{a}(\alpha_a,\beta_a)\equiv g^{-1}(\alpha_a,\beta_a),~~\\
     & \forall X^a,Y^a\in\Sigma^a ,~~~~
      \forall \alpha_a,\beta_a\in\Sigma^{*}_a~.   \nonumber
\end{eqnarray}

In these relations $\{g^a\}$ are the metrics induced on the subbundles
$\{\Sigma^a\}$ of the tangent bundle $T(M)$. Using this scheme we can obtain
the decomposition of more complex tensors. We assume that all objects
with indices $a,b,...$ are defined on the associated subbundles
$\Sigma^a,~\Sigma^b,...$.

Let $\nabla$ be an affine (symmetric and compatible with $g$) connection such
that
\begin{equation}\label{11}
\nabla_{X}Y - \nabla_{Y}X = [X,Y],~~
X(Y\cdot Z) = Z\cdot\nabla_{X}Y + Y\cdot\nabla_{X}Z
\end{equation}
where $[X,Y]=XY-YX$ is the Lie bracket of two vector fields $X$ and $Y$,
$\nabla_{X}Y$ is the covariant derivative of $Y$ in the direction $X$.
A consequence of this is that
\begin{equation}\label{12}
    2Z\cdot\nabla_{X}Y = X(Y\cdot Z)+Y(Z\cdot X)-Z(X\cdot Y)+
Z\cdot [X,Y] + Y\cdot [Z,X] - X\cdot [Y,Z].
\end{equation}

Then the covariant derivative $\nabla_{X}T$ of a tensor $T$ of type
$(s,r)$, where $s=0,1$ with respect to $X$ is defined by
\begin{equation}\label{13}
   (\nabla_{X}T)(Y_1,...Y_r) = \nabla_{X}(T(Y_1,...Y_r)) -
\sum^{r}_{i=1}T(Y_{1},...Y_{i-1},\nabla_{X}Y_{i},Y_{i+1},...Y_{r}).
\end{equation}
The Lie derivative ${\cal L}_{X}T$ of a tensor $T$ with respect to a
vector $X$ and the exterior derivative of an $r$-form $\Omega$ are given by:
\begin{eqnarray}\label{14}
   ({\cal L}_{X}T)(Y_1,...Y_r) &=& {\cal L}_{X}(T(Y_1,...Y_r))-
\sum^{r}_{i=1}T(Y_{1},...Y_{i-1},{\cal L}_{X}Y_{i},Y_{i+1},...Y_{r})
\nonumber \\
   (d\Omega)(Y_0,Y_1,...Y_r) &=&
\sum^{r}_{i=0}(-1)^i Y_i(\Omega(Y_{0},...,\hat Y_{i},...Y_{r})) \\
 &+&  \sum_{0\leq i<j\leq r}(-1)^{i+j} \Omega([Y_{i},Y_{j}],Y_{0},...,
\hat Y_{i},...,\hat Y_{j},...,Y_{r}) \nonumber
\end{eqnarray}
where ${\cal L}_{X}Y = [X,Y],~~
{\cal L}_{X}\varphi = \nabla_{X}\varphi = (d\varphi)(X) = X\varphi$
for any scalar function $\varphi$; the symbol $~^{\wedge}$ means that the
associated term is omitted.

The curvature tensor is defined by the formula
\begin{equation}\label{15}
R(X,Y)Z = (\nabla_{X}\nabla_{Y}-\nabla_{Y}\nabla_{X}-\nabla_{[X,Y]})Z.
\end{equation}

Using a split structure ${\cal H}^r$, the decomposition of
$\nabla$ is easily set up:
\begin{equation}\label{16}
\nabla_{X}Y = \sum^{r}_{a,b,c=1}\nabla^{a}_{X^b}Y^{c},~~~~~
\forall X,Y\in T(M).
\end{equation}
In this sum we can distinguish the five different sorts of objects
$\{\nabla^{a}_{X^a}Y^{a},~~\nabla^{a}_{X^b}Y^{b},~~ \nabla^{a}_{X^a}Y^{b}, \\
 \nabla^{a}_{X^b}Y^{a}~~\nabla^{a}_{X^b}Y^{c}\}~~(a\not=b\not=c)$, which
complete the whole of the projected connections. In particular, in
this sum the objects
\begin{equation}\label{17}
\nabla^{a}_{X^a}Y^{a}\equiv H^{a}\cdot\nabla_{X^a}Y^{a},~~~~~~
\forall X^{a},Y^{a}\in\Sigma^a~~~~(a=1,2,...r)
\end{equation}
define connections $\{\nabla^a\}$ induced on the subbundles $\{\Sigma^a\}$.
The object
\begin{equation}\label{18}
\nabla^{a}_{X^b}Y^{b}\equiv H^{a}\cdot\nabla_{X^b}Y^{b}\equiv
-B^{a}(X^{b},Y^b),~~~~~
\forall X^{b},Y^{b}\in\Sigma^b
\end{equation}
is the tensor of extrinsic non-holonomicity of the subbundle $\Sigma^b$.
One can think that the objects
\begin{equation}\label{19}
\nabla^{a}_{X^b}Y^{c}\equiv H^{a}\cdot\nabla_{X^b}Y^{c}\equiv
- Q^{a}(X^b,Y^c)\equiv - Q^{a}_{bc}(X^b,Y^c),~~~~~\forall a\not=b\not=c
\end{equation}
define the generalization of the Ricci's coefficients of rotation
$\gamma^{a}_{bc}$~\cite{Land}. In general
case they give the objects of rotation $Q^{a}_{bc}$ of the subbundles
$\Sigma^b,~\Sigma^c$ in the $n_a$-dimensional direction $\Sigma^a$.
The other components can be expressed in terms of the introduced objects and
the Lie derivative ${\cal L}_{X^b}Y^c$ projected
into every possible subbundle $\Sigma^a~~(a\not=b\not=c)$. Thus, the components
$\nabla^{a}_{X^a}Y^{b}\equiv H^{a}\cdot\nabla_{X^a}Y^{b}$ and
$\nabla^{a}_{X^b}Y^{a}\equiv H^{a}\cdot\nabla_{X^b}Y^{a}$ satisfy the
relations
\begin{equation}\label{20}
Z^{a}\cdot\nabla^{a}_{X^a}Y^{b} = Y^{b}\cdot B^{b}(X^a,Z^a )~~~~
      (a\not=b)
\end{equation}
\begin{equation}\label{21}
Z^{a}\cdot\nabla^{a}_{X^b}Y^{a} = Z^{a}\cdot\Lambda^{a}(X^b,Y^a)+
X^{b}\cdot B^{b}(Y^a,Z^a)
\end{equation}
where
\begin{equation}\label{22}
\Lambda^{a}(X^b,Y^c) = [X^{b},Y^{c}]^{a}\equiv H^{a}\cdot [X^{b},Y^{c}]
     \equiv {\cal L}^{a}_{X^b}Y^c ~~~~(a\not=b\not=c).
\end{equation}
Taking into account the relation (\ref{12}) and the definition (\ref{19})
we have
$$    2Z^{a}\cdot Q^{a}(X^b,Y^c)=X^{b}\cdot\Lambda^{b}(Y^{c},Z^a)
    -Z^{a}\cdot\Lambda^{a}(X^b,Y^c)-Y^{c}\cdot\Lambda^{c}(Z^a,X^b). $$

The tensor of extrinsic non-holonomicity can be expressed as the sum of
symmetric and antisymmetric parts
\begin{equation}\label{23}
B^{a}(X^{b},Y^b ) = S^{a}(X^{b},Y^b ) + A^{a}(X^{b},Y^b )
\end{equation}
where $S^{a}(X^{b},Y^b )$ and $A^{a}(X^{b},Y^b )$ define the tensors of
extrinsic curvature and extrinsic torsion of subbundle $\Sigma^b$ in the
direction of the subbundle $\Sigma^a$. For these objects one can obtain the
relations
\begin{equation}\label{24}
2Z^{a}\cdot S^{a}(X^{b},Y^b ) = ({\cal L}_{Z^{a}}g^b)(X^{b},Y^b )
\end{equation}
\begin{equation}\label{25}
2A^{a}(X^{b},Y^b ) = -[X^{b},Y^{b}]^{a}\equiv
-H^{a}\cdot [X^{b},Y^{b}].
\end{equation}
It easy can be shown that a connection $\nabla^a$ induced on the subbundle
$\Sigma^a$ will be symmetric and compatible with the metric $g^a$. The
projecting of the curvature tensor into the subbundles
$\Sigma^{a},\Sigma^{b},...$ gives us the nonholonomic generalizations of the
Gauss-Codazzi-Ricci's equations. Using the definitions  (\ref{8}-\ref{10}),
(\ref{15}-\ref{25}) one can obtain all the necessary projections of the
curvature tensor (for more details see Appendix A).

\section{Invariant \ab ~Decomposition of a Pseudo-Riemannian manifold}

If $r=2$, then there are only two subbundles $\Sigma^\prime$ and
$\Sigma^{\prime\prime}$ of the tangent bundle $T(M)$ and the previous
formulae become much simpler. Owing to importance of this case it was deemed
worthwhile to consider the split structure independently from SecII
~\cite{Gl3}, \cite{Gl4}.

Let $H^\prime$ be a linear idempotent symmetric operator of a constant rank
with the property
\begin{equation}\label{26}
H^\prime\cdot H^\prime = H^\prime.
\end{equation}

We shall say that $H^\prime$ defines {\bf a $(n+m)$ split structure} on $M$ if
\begin{equation}\label{27}
{\rm dim}~{\rm Im} H^\prime = n;~~
{\rm dim}~{\rm Ker} H^\prime = m;~~
{\rm dim} ~M = n+m
\end{equation}
where ${\rm Ker} H^\prime$ is a kernel of the operator $H^\prime$.
Since $H^{\prime}$ is defined, thereby we define the operator
$H^{\prime\prime}$ such that
\begin{equation}\label{28}
H^{\prime\prime}\cdot H^{\prime\prime} = H^{\prime\prime};~~
H^{\prime\prime}\cdot H^{\prime} = H^{\prime}\cdot H^{\prime\prime} = 0;~~
H^{\prime} + H^{\prime\prime} = I.
\end{equation}
Therefore $H^{\prime}$ and $H^{\prime\prime}$ are the projection operators
which determine the split structure
${\cal H}^2$ on $M$ due to the definition (\ref{4}). We have
\begin{eqnarray}\label{29}
    T(M)=\Sigma^{\prime}\oplus\Sigma^{\prime\prime};~~
\Sigma^{\prime} = {\bigcup_{p\in M}}\Sigma^{\prime}_p;~~
\Sigma^{\prime}_p = {\rm Im} H^{\prime}_p;~~
\Sigma^{\prime\prime} = {\bigcup_{p\in M}}\Sigma^{\prime\prime}_p;~~
\Sigma^{\prime\prime}_p = {\rm Ker}~H^{\prime}_p;
\end{eqnarray}
\begin{eqnarray}\label{30}
X = X^{\prime} + X^{\prime\prime};~~~
g = g^{\prime} + g^{\prime\prime};~~~
g^{-1} = (g^{\prime})^{-1} + (g^{\prime\prime})^{-1}; \nonumber  \\
X^{\prime} = H^{\prime}\cdot X;~~~
X^{\prime\prime} = H^{\prime\prime}\cdot X;~~~
X^{\prime}\cdot X^{\prime\prime} = 0;                         \\
g^{\prime}(X^{\prime},Y^{\prime}) = g(X^{\prime},Y^{\prime});~~~
g^{\prime\prime}(X^{\prime\prime},Y^{\prime\prime}) =
g(X^{\prime\prime},Y^{\prime\prime}). \nonumber
\end{eqnarray}
A connection $\nabla$ is decomposed into the following components:
a connection on $\Sigma^\prime$, and the tensor of extrinsic non-holonomicity
of the subbundle $\Sigma^\prime$, respectively
\begin{equation}\label{33}
\nabla^{\prime}_{X^{\prime}}Y^\prime =
H^{\prime}\cdot\nabla_{X^{\prime}}Y^\prime
\end{equation}
\begin{equation}\label{34}
B^{\prime\prime}(X^{\prime},Y^{\prime}) =
- \nabla^{\prime\prime}_{X^{\prime}}Y^\prime =
- H^{\prime\prime}\cdot\nabla_{X^{\prime}}Y^\prime .
\end{equation}
Other components of $\nabla$ can be expressed in terms of the components
(\ref{33}),(\ref{34}) and the Lie derivatives of two vector fields
\begin{equation}\label{35}
X^{\prime}\cdot\nabla^{\prime}_{Y^\prime}Z^{\prime\prime} =
Z^{\prime\prime}\cdot B^{\prime\prime}(Y^{\prime},X^{\prime})
\end{equation}
\begin{equation}\label{36}
X^{\prime}\cdot\nabla^{\prime}_{Y^{\prime\prime}}Z^{\prime} =
X^{\prime}\cdot {\cal L}_{Y^{\prime\prime}} Z^{\prime} +
Y^{\prime\prime}\cdot B^{\prime\prime}(Z^{\prime},X^{\prime}).
\end{equation}
The rest of the components of $\nabla$
$\{\nabla^{\prime\prime}_{X^{\prime\prime}}Y^{\prime\prime},~~
\nabla^{\prime}_{X^{\prime\prime}}Y^{\prime\prime},~~
\nabla^{\prime\prime}_{X^{\prime\prime}}Y^\prime,~~
\nabla^{\prime\prime}_{X^{\prime}}Y^{\prime\prime}\}$
may be written out by substituting
$X^{\prime},Y^{\prime},B^{\prime},H^{\prime},...$ for
$X^{\prime\prime},Y^{\prime\prime},B^{\prime\prime},H^{\prime\prime},...$ and
vice versa in formulae (\ref{33})-(\ref{36}). This completes the set of all
the eight possible projections of the connection.

The tensor $B^{\prime\prime}$ may be expressed as the sum of its symmetric and
antisymmetric parts:
\begin{equation}\label{37}
B^{\prime\prime}(X^{\prime},Y^\prime ) =
S^{\prime\prime}(X^{\prime},Y^\prime ) + A^{\prime\prime}(X^{\prime},Y^\prime )
\end{equation}
\begin{equation}\label{38}
    2Z^{\prime\prime}\cdot S^{\prime\prime}(X^{\prime},Y^\prime ) =
({\cal L}_{Z^{\prime\prime}}g^{\prime})(X^{\prime},Y^{\prime});~~~~~
2A^{\prime\prime}(X^{\prime},Y^\prime ) =
-H^{\prime\prime}\cdot [X^{\prime},Y^{\prime}]
\end{equation}
where $S^{\prime\prime}$ and $A^{\prime\prime}$ are the tensors of extrinsic
curvature and torsion respectively. If $A^{\prime\prime}=0$, the subbundle
$\Sigma^\prime$ will be holonomic.
It means that the subbundle $\Sigma^\prime$ is the union of the tangent
bundles of an $m$-parameter family of $n$-dimensional surfaces
$\{M^n (q)\subset M\}$, where $q=\{c^i\}\in D \subset R^m$ parameterizes the
surfaces $M^n (q)$, and $D$ is some range of parameters $c^i~(i=1,2,...,m)$
in $R^m$, that is $\Sigma^\prime={\bigcup_{q\in D}} T(M^n(q))$. This
implies that a covector basis of locally exact one-forms $\{dx^i\}$ exists
on the dual subbundle $(\Sigma^{\prime\prime})^{*}$, so that each of the surfaces of
$\{M^n (q)\}$ is the
intersection of hypersurfaces $x^i=c^i~(i=1,2,...,m)$ for some values
of $c^i \in D$.

Using the definition of the curvature tensor (\ref{15}) one can find every possible
projection of the curvature tensor
\begin{eqnarray} \label{39}
    R(X^{\prime},Y^{\prime})Z^{\prime}\cdot V^{\prime} &=&
R^{\prime}(X^{\prime},Y^{\prime})Z^{\prime}\cdot V^{\prime} +
B^{\prime\prime}(X^{\prime},Z^{\prime})\cdot
B^{\prime\prime}(Y^{\prime},V^{\prime})  \nonumber \\
&-& ~B^{\prime\prime}(Y^{\prime},Z^{\prime})\cdot
B^{\prime\prime}(X^{\prime},V^{\prime}) +
2A^{\prime\prime}(X^{\prime},Y^{\prime})\cdot
B^{\prime\prime}(Z^{\prime},V^{\prime});
\end{eqnarray}
\begin{eqnarray} \label{40}
    R(X^{\prime},Y^{\prime})Z^{\prime}\cdot V^{\prime\prime} &=&
V^{\prime\prime}\cdot \{(\nabla^{\prime\prime}_{Y^{\prime}} B^{\prime\prime})
(X^{\prime},Z^{\prime})-
(\nabla^{\prime\prime}_{X^{\prime}} B^{\prime\prime})
(Y^{\prime},Z^{\prime})\} \nonumber \\
&+& ~2Z^{\prime}\cdot
B^{\prime}(A^{\prime\prime}(X^{\prime},Y^{\prime}),V^{\prime\prime});
\end{eqnarray}
\begin{eqnarray} \label{41}
    R(X^{\prime},Y^{\prime\prime})Z^{\prime}\cdot V^{\prime\prime} &=&
(Z^{\prime}\cdot (\nabla^{\prime}_{X^{\prime}} B^{\prime})+
<X^{\prime}\cdot B^{\prime},Z^{\prime}\cdot B^{\prime}>)
(Y^{\prime\prime},V^{\prime\prime})                              \nonumber \\
&+& ~(V^{\prime\prime}\cdot(\nabla^{\prime\prime}_{Y^{\prime\prime}}
B^{\prime\prime})+
<Y^{\prime\prime}\cdot B^{\prime\prime}, V^{\prime\prime}\cdot
B^{\prime\prime}>)
(X^{\prime},Z^{\prime});
\end{eqnarray}
where $R^{\prime}$ is the curvature tensor of the subbundle
$\Sigma^{\prime}$
\begin{equation}\label{42}
    R^{\prime}(X^{\prime},Y^{\prime})Z^\prime \equiv
\{\nabla^{\prime}_{X^\prime}\nabla^{\prime}_{Y^\prime} -
\nabla^{\prime}_{Y^\prime}\nabla^{\prime}_{X^\prime} -
\nabla^{\prime}_{[X^{\prime},Y^{\prime}]^{\prime}} +
2{\cal L}^{\prime}_{A^{\prime\prime}(X^{\prime},Y^{\prime})}\}Z^{\prime},~~
\forall X^{\prime},Y^{\prime},Z^\prime\in\Sigma^{\prime}
\end{equation}
where ${\cal L}^{\prime}$ - the Lie derivative projected into the subbundle
$\Sigma^{\prime}~~({\cal L}^{\prime}_{X}Y\equiv H^{\prime}\cdot{\cal L}_{X}Y)$.
This definition of the curvature tensor, introduced in the works
\cite{Gl2}-\cite{Gl4}, is the invariant generalization of
that introduced in coordinate form in \cite{Zel1}-\cite{Zel3} by analogy with
\cite{Schout}. Note that the latter term in (\ref{42}) is necessary in order
that the differential curvature operator
$R^{\prime}(X^{\prime},Y^{\prime})$ on $\Sigma^{\prime}$ be a linear
multiplicative one, or, in other words, $R^{\prime}$
be a tensor of type (1,3) on non-holonomic subbundle $\Sigma^{\prime}$.
In similar fashion this concerns the general
case of ${\cal H}^r$ split structure (see Appendix A, (\ref{A6}) for
$R^{a}(X^{a},Y^{a})Z^a$).

The following expression, with the fixed vectors
$X^{\prime},Z^{\prime},Y^{\prime\prime},V^{\prime\prime}$,
$$ (<Y^{\prime\prime}\cdot B^{\prime\prime}, V^{\prime\prime}\cdot
B^{\prime\prime}>)(X^{\prime},Z^{\prime})\equiv
<Y^{\prime\prime}\cdot B^{\prime\prime}(X^{\prime},~.~),
 V^{\prime\prime}\cdot B^{\prime\prime}(~.~,Z^{\prime})> $$
defines  the  scalar  product  of the two one-forms
$\alpha\equiv Y^{\prime\prime}\cdot B^{\prime\prime}(X^{\prime},~.~)$ and
$\beta\equiv V^{\prime\prime}\cdot B^{\prime\prime}(~.~,Z^{\prime})$
according to (\ref{1}) by  the metric $(g^{\prime})^{-1}$.
The covariant derivatives of the tensor $B^{\prime}$ are given by
\begin{equation}
   (\nabla^{\prime}_{X^{\prime}} B^{\prime})(Y^{\prime\prime},Z^{\prime\prime})
= \nabla^{\prime}_{X^{\prime}} (B^{\prime}(Y^{\prime\prime},Z^{\prime\prime}))-
B^{\prime}
(\nabla^{\prime\prime}_{X^{\prime}}Y^{\prime\prime},Z^{\prime\prime})-
B^{\prime}
(Y^{\prime\prime},\nabla^{\prime\prime}_{X^{\prime}}Z^{\prime\prime})
\label{43}
\end{equation}
\begin{equation}
   (\nabla^{\prime}_{X^{\prime\prime}}B^{\prime})
(Y^{\prime\prime},Z^{\prime\prime}) =
\nabla^{\prime}_{X^{\prime\prime}}(B^{\prime}
(Y^{\prime\prime},Z^{\prime\prime})) -
B^{\prime}
(\nabla^{\prime\prime}_{X^{\prime\prime}}
Y^{\prime\prime},Z^{\prime\prime}) -
B^{\prime}
(Y^{\prime\prime},\nabla^{\prime\prime}_{X^{\prime\prime}}Z^{\prime\prime}).
\label{44}
\end{equation}

The relations (\ref{39})-(\ref{41}) are nonholonomic analogies of the
well-known Gauss-Codazzi-Ricci's equations. Other nontrivial
projections of the curvature tensor may be written out using the
substitution "$~^{\prime}~$" for "$~^{\prime\prime}~$" and vice versa.

In the special case of coordinate representation of $(3+1)$ and $(2+2)$
decomposition, the objects introduced above give us the known tensors
~\cite{Zel1}~-\cite{Pol}, which have clear physical and geometrical meaning.

Let us note that the objects, presented in the work ~\cite{Gray} may be expressed in
terms of these tensors. For example, the torsion tensor introduced there as
the Nijenhuis tensor ~\cite{Kob} proved to be equal
\begin{eqnarray}
  S_{H^{\prime}}(X,Y)\equiv
[X,Y^{\prime}]^\prime + [X^{\prime},Y]^\prime -
[X^{\prime},Y^{\prime}] - [X,Y]^\prime =
2A^{\prime}(X^{\prime\prime},Y^{\prime\prime}) +
2A^{\prime\prime}(X^{\prime},Y^\prime )~ \nonumber
\end{eqnarray}
and the tensors $T_{X}Y$ and $Q_{X}Y$of the work ~\cite{Gray} are given by:
\begin{eqnarray}
T_{X}Y\equiv
\nabla^{\prime\prime}_{X^\prime}Y^\prime +
\nabla^{\prime}_{X^\prime}Y^{\prime\prime} =
-B^{\prime\prime}(X^{\prime},Y^{\prime}) +
g^{-1}(Y^{\prime\prime}\cdot B^{\prime\prime}(X^{\prime},~.~),~.~) \nonumber\\
Q_{X}Y\equiv
\nabla^{\prime}_{X^{\prime\prime}}Y^{\prime\prime} +
\nabla^{\prime\prime}_{X^{\prime\prime}}Y^\prime =
-B^{\prime}(X^{\prime\prime},Y^{\prime\prime}) +
g^{-1}(Y^{\prime}\cdot B^{\prime}(X^{\prime\prime},~.~),~.~). \nonumber
\end{eqnarray}
They have not any simple interpretation even in the classical case of
hypersurfaces in $M$.

\section{An invariant \cd ~Split structure on a
          Pseudo-Riemannian manifold}

In this section we give the invariant generalization of $(n+1)$
decomposition of spaces (the monad method ~\cite{Zel3},~\cite{Vlad2}) as
a special case of  $(n+m)$ decomposition when $m=1$.

Let $u$ be a vector field (field of a monad) on $M$ such that
$u\cdot u=\varepsilon = \pm 1$. It gives a one-form $\omega$ and projection
operators uniquely by the formulae
\begin{equation}\label{45}
\omega(X) = \varepsilon u\cdot X,~~~~~~\forall X\in T(M)
\end{equation}
\begin{equation}\label{46}
H^{\prime\prime} = u\otimes\omega;~~~~~~
H^{\prime} = I-H^{\prime\prime}.
\end{equation}
The operators satisfy all the necessary relations (\ref{26})-(\ref{28})
when $\Sigma^{\prime\prime}$ is a one-dimensional subbundle $(m=1)$.
The tensor product is denoted by "$\otimes$".

Thus, defining vector or covector fields, $u$ or $\omega$ respectively, we,
thereby, induce an $(n+1)$ split structure on $M$. For any vector field $X$ and
metric $g$, this implies
\begin{equation}\label{47}
X = X^{\prime} + \omega (X)u ,~~~
g = g^{\prime} + \varepsilon\omega\otimes\omega ,~~~
g^{-1} = (g^{\prime})^{-1} + \varepsilon u\otimes u~.
\end{equation}
Hence it is apparent that $X^{\prime\prime}=\omega (X)u$ is collinear to $u$.
The metrics $g^{\prime\prime}=\varepsilon\omega\otimes\omega$,
$~(g^{\prime\prime})^{-1}=\varepsilon u\otimes u$ and $g^{\prime}$,
$(g^\prime)^{-1}$ are the metrics on the subbundles $\Sigma^{\prime\prime}$,
$\Sigma^{*\prime\prime}$ and $\Sigma^\prime$, $\Sigma^{*\prime}$,
correspondingly. A connection $\nabla$ has the following components:
\begin{equation}\label{48}
\nabla_{X^\prime}Y^\prime =
\nabla^{\prime}_{X^\prime}Y^\prime -
B(X^{\prime},Y^{\prime})u;~~~~
\nabla_{u}u = \nabla^{\prime}_{u}u =
-B^{\prime}(u,u)\equiv F
\end{equation}
where $B(X^{\prime},Y^{\prime})=
\omega(B^{\prime\prime}(X^{\prime},Y^{\prime}))$. The latter equation in
(\ref{48}) follows from the formula $u\cdot u=\varepsilon = \pm 1$.
If we consider a congruence of curves for which the vector $u$ is the tangent
vector, then $F$ is the first curvature of this congruence.
The tensor $B$ of type (0,2) is the tensor of extrinsic non-holonomicity
of the subbundle $\Sigma^\prime$ and can be written as the sum of its
symmetric and antisymmetric parts:
\begin{equation}\label{49}
B(X^{\prime},Y^\prime ) =
-\omega (\nabla^{\prime\prime}_{X^\prime}Y^\prime ) =
\varepsilon S(X^{\prime},Y^\prime ) + A(X^{\prime},Y^\prime ),
\end{equation}
where   $S(X^{\prime},Y^\prime )=
\varepsilon\omega(S^{\prime\prime}(X^{\prime},Y^\prime )),~~~
A(X^{\prime},Y^\prime )=\omega(A^{\prime\prime}(X^{\prime},Y^\prime ))$
and
\begin{equation}\label{50}
2S(X^{\prime},Y^\prime ) =
({\cal L}_{u}g^{\prime})(X^{\prime},Y^{\prime});~~~~~
2A(X^{\prime},Y^\prime ) =
(d\omega)(X^{\prime},Y^{\prime})
\end{equation}
are the tensors of extrinsic curvature and extrinsic torsion of the subbundle
$\Sigma^\prime$.

The components of the curvature tensor in an $(n+1)$ decomposed form lead
to the generalized Gauss-Codazzi-Ricci's equations:
\begin{eqnarray} \label{51}
    R(X^{\prime},Y^{\prime})Z^{\prime}\cdot V^{\prime} &=&
R^{\prime}(X^{\prime},Y^{\prime})Z^{\prime}\cdot V^{\prime} +
\varepsilon [2A(X^{\prime},Y^{\prime})B(Z^{\prime},V^{\prime})  \nonumber \\
      &+&  ~B(X^{\prime},Z^{\prime})B(Y^{\prime},V^{\prime}) -
B(Y^{\prime},Z^{\prime})(X^{\prime},V^{\prime})]
\end{eqnarray}
\begin{equation}\label{52}
    R(X^{\prime},Y^{\prime})Z^{\prime}\cdot u =
-2A(X^{\prime},Y^{\prime})F\cdot Z^\prime +
\varepsilon [(\nabla_{Y^\prime} B)(X^{\prime},Z^{\prime}) -
(\nabla_{X^\prime} B)(Y^{\prime},Z^{\prime})]
\end{equation}
\begin{equation}\label{53}
    R(X^{\prime},u)Y^{\prime}\cdot u =
- Y^{\prime}\cdot\nabla^{\prime}_{X^\prime}F +
\varepsilon (F\cdot X^{\prime})(F\cdot Y^{\prime}) +
(\varepsilon {\cal L}_{u}B -
<B,B^{T}>)(X^{\prime},Y^{\prime})~
\end{equation}
where the curvature tensor of the subbundle $\Sigma^\prime$ (see \cite{Gl1})
is given by
\begin{eqnarray}
    R^{\prime}(X^{\prime},Y^{\prime})Z^\prime \equiv
\{\nabla^{\prime}_{X^\prime}\nabla^{\prime}_{Y^\prime} -
\nabla^{\prime}_{Y^\prime}\nabla^{\prime}_{X^\prime} -
\nabla^{\prime}_{[X^\prime ,Y^\prime ]^\prime} +
2A(X^\prime ,Y^\prime ){\cal L}^{\prime}_{u}\}Z^{\prime}~.
\end{eqnarray}
It is to be noted that for tensors of an arbitrary type
the projection operators are constructed by the tensor product of the
operators (\ref{46}) and their transposes. If one does no more than
$(n+1)$ decomposition of objects only from the Cartan
algebra of exterior forms on $M$ then the universal invariant construction
of the projection operators is feasible (see for example \cite{Fecko} for
$(3+1)$ decomposition).

\section {\ab ~Decomposition of a Pseudo-Riemannian manifold in
          coordinate form}

In order to obtain a coordinate form of the invariant objects it is necessary
to choose coordinate covector and vector bases
$\{\partial_{\mu}=\partial / \partial x^\mu\},~\{dx^\mu\}$ in the
domain ${\rm U}$ of some map $x^\mu~(\mu, \nu, \rho,... =1,2,...n,n+1,...n+m)$.
Then we can find all the relations given above with respect to this basis,
i.e. in covariant form.

Thus in the case of an $(n+m)$ decomposition one has
\begin{eqnarray}\label{54}
   & H^\prime = h^{\prime\nu}_{~\mu}\partial_{\nu}\otimes dx^\mu
  =h^{\prime\nu}_{~\mu}\partial^{\prime}_{\nu}\otimes d^{\prime}x^\mu;~~~
  \partial^{\prime}_{\mu}\equiv h^{\prime\nu}_{~\mu}\partial_\nu,~~~
  d^{\prime}x^{\mu}\equiv h^{\prime\mu}_{~\nu}dx^\nu \nonumber \\
  & H^{\prime\prime} = h^{\prime\prime\nu}_{~\mu}\partial_{\nu}\otimes dx^\mu
 =h^{\prime\prime\nu}_{~\mu}\partial^{\prime\prime}_{\nu}\otimes
      d^{\prime\prime}x^\mu;~~~
 \partial^{\prime\prime}_{\mu}\equiv h^{\prime\prime\nu}_{~\mu}\partial_\nu,~~~
  d^{\prime\prime}x^{\mu}\equiv h^{\prime\prime\mu}_{~\nu}dx^\nu  \\
  & h^{\prime\nu}_{~\mu}h^{\prime\mu}_{~\rho}=h^{\prime\nu}_{~\rho};~~~
   h^{\prime\prime\nu}_{~\mu}h^{\prime\prime\mu}_{~\rho}=
   h^{\prime\prime\nu}_{~\rho};~~~
   h^{\prime\nu}_{~\mu}h^{\prime\prime\mu}_{~\rho}=0;~~~
   h^{\prime\nu}_{~\mu}+h^{\prime\prime\nu}_{~\mu}=\delta^{\nu}_{\mu} \nonumber
\end{eqnarray}
\begin{eqnarray}\label{55}
    & g=g^{\prime}+g^{\prime\prime}=
  g^{\prime}_{\mu\nu} d^{\prime}x^{\mu}\otimes d^{\prime}x^{\nu}+
  g^{\prime\prime}_{\mu\nu} d^{\prime\prime}x^{\mu}
  \otimes d^{\prime\prime}x^{\nu}   \nonumber \\
   & g_{\mu\nu}\equiv \partial_\mu\cdot\partial_\nu =
  g^{\prime}_{\mu\nu} + g^{\prime\prime}_{\mu\nu};~~~
  g^{\prime}_{\mu\nu}=
   h^{\prime\rho}_{~\mu}h^{\prime\sigma}_{~\nu}g_{\rho\sigma}~~~
  g^{\prime\prime}_{\mu\nu}=
   h^{\prime\prime\rho}_{~\mu}h^{\prime\prime\sigma}_{~\nu}g_{\rho\sigma}
\end{eqnarray}
Further, introducing the definitions
\begin{equation}\label{56}
    [\partial^{\prime}_{\mu},\partial^{\prime}_{\nu}]^\prime \equiv
\lambda^{\rho^\prime}_{\mu^{\prime}\nu^\prime}\partial^{\prime}_\rho;~~~
[\partial^{\prime}_{\mu},\partial^{\prime\prime}_{\nu}]^\prime\equiv
\lambda^{\rho^\prime}_{\mu^{\prime}\nu^{\prime\prime}}
\partial^{\prime}_\rho;~~~
[\partial^{\prime\prime}_{\mu},\partial^{\prime\prime}_{\nu}]^{\prime}\equiv
-2A^{\rho^\prime}_{\mu^{\prime\prime}\nu^{\prime\prime}}\partial^{\prime}_\rho
\end{equation}
\begin{equation}\label{57}
\nabla^{\prime}_{\partial^{\prime}_\mu}\partial^{\prime}_\nu\equiv
L^{\rho^{\prime}}_{\mu^{\prime}\nu^{\prime}} \partial^{\prime}_{\rho};~~~~
 B^\prime\equiv
B^{\rho^{\prime}}_{\mu^{\prime\prime}\nu^{\prime\prime} } \partial^{\prime}_\rho
\otimes d^{\prime\prime}x^{\mu}\otimes d^{\prime\prime}x^{\nu};~~~~
\end{equation}
one has
\begin{equation}\label{58}
L^{\rho^\prime}_{\mu^{\prime}\nu^\prime} =
d^{\prime}x^{\rho}(\nabla_{\partial^{\prime}_\mu} \partial^{\prime}_\nu);~~~~~
B^{\rho^\prime}_{\mu^{\prime\prime}\nu^{\prime\prime}} =
dx^{\prime\rho}(\nabla_{\partial^{\prime\prime}_\mu}\partial^{\prime\prime}_\nu ) =
S^{\rho^{\prime}}_{\mu^{\prime\prime}\nu^{\prime\prime}} +
A^{\rho^\prime}_{\mu^{\prime\prime}\nu^{\prime\prime}}
\end{equation}
\begin{eqnarray} \label{59}
    2A^{\rho^\prime}_{\mu^{\prime\prime}\nu^{\prime\prime}} =
h^{\prime\prime\omega}_{~\mu}h^{\prime\prime\gamma}_{~\nu}
(h^{\prime\rho}_{~\gamma,\omega} - h^{\prime\rho}_{~\omega,\gamma});~~~~
2S_{\rho^\prime \mu^{\prime\prime}\nu^{\prime\prime}} =
\partial^{\prime}_\rho g^{\prime\prime}_{\mu\nu} +
g^{\prime\prime}_{\mu\sigma}
\lambda^{\sigma^{\prime\prime}}_{\nu^{\prime\prime}\rho^\prime} +
g^{\prime\prime}_{\sigma\nu}
\lambda^{\sigma^{\prime\prime}}_{\mu^{\prime\prime}\rho^\prime}.
\end{eqnarray}
Here $h_{,\gamma}\equiv\partial h/\partial x^{\gamma};~~
\mu^{\prime},\nu^{\prime},\rho^{\prime},...,
\mu^{\prime\prime},\nu^{\prime\prime},\rho^{\prime\prime},...
=1,2,...n,n+1,...n+m$. The indices "$~^{\prime}~$" and "$~^{\prime\prime}~$"
indicate that the corresponding objects are associated with the subbundles
$\Sigma^{\prime}$ and $\Sigma^{\prime\prime}$ respectively. From the
previous formulae it follows that there are the objects which are associated
with the both subbundles $\Sigma^{\prime}$ and $\Sigma^{\prime\prime}$.
For instance, the tensor of extrinsic non-holonomicity
$B^{\rho^\prime}_{\mu^{\prime\prime}\nu^{\prime\prime}}$ is a
contravariant vector on the subbundle $\Sigma^{\prime}$, and a covariant
tensor of rank 2 on the subbundle $\Sigma^{\prime\prime}$.

The other necessary objects can be found by substituting
"$~^{\prime}~$" for "$~^{\prime\prime}~$" and vice versa.
Using these formulae we can obtain
the Gauss-Codazzi-Ricci's equations in terms of the introduced objects. If
$n=m=2$, our treatment is reduced to
the dyad formalism (see \cite{Pol}).

In the case of an $(n+1)$ split structure (see SecIV), we have
$u= u^\mu \partial_\mu ,~(\mu,\nu =1,2,...,n,n+1)$, and

\begin{eqnarray}
      u_\mu u^\mu = \varepsilon &=& \pm 1,~~~~
h^{\prime\prime~\nu}_{~\mu} = \varepsilon u_\mu u^\nu,~~~~
h^{\prime~\nu}_{~\mu} = \delta_{\mu}^{~\nu}
      - \varepsilon u_\mu u^\nu              \nonumber \\
g_{\mu\nu} &=& \varepsilon u_\mu u_\nu + g^{\prime}_{\mu\nu};~~~~~~
g^{\mu\nu} = \varepsilon u^\mu u^\nu + g^{\prime\mu\nu} \\
g^\prime_{\mu\nu} &=& h^{\prime~\alpha}_{~\mu}h^{\prime~\beta}_{~\nu}
      g_{\alpha\beta};~~~~~~
g^{\prime\mu\nu} = h^{\prime\mu}_{~\alpha}h^{\prime\nu}_{~\beta}
g^{\alpha\beta}
\label{60}
\end{eqnarray}
\begin{eqnarray} \label{61}
\partial^{\prime}_\mu = h^{\prime\nu}_{~\mu}\partial_\nu ;~~~~~
[\partial^{\prime}_{\mu},\partial^{\prime}_{\nu}] =
\varepsilon A_{\mu\nu} u ;~~~~~
[\partial^{\prime}_{\mu},u] = -F_\mu u
\end{eqnarray}
\begin{equation}\label{62}
    2A_{\mu\nu} = h^{\prime\rho}_{,\mu}h^{\prime\sigma}_{,\nu}
(u_{\rho,\sigma} - u_{\sigma,\rho});~~~~
F_\mu = (u_{\mu,\nu} - u_{\nu,\mu})u^\nu ;~~~~
2S_{\mu\nu} = {\cal L}_u g^\prime_{\mu\nu}.
\end{equation}
Replacing all the objects in (\ref{51})-(\ref{53}) by these relations we can find the
components of the curvature tensor. Furthermore if we consider $(3+1)$ decomposition
of a relativistic space-time, our formalism is reduced to the monad method
~\cite{Zel1}-~\cite{Vlad2}, and to his special gauges. In this case abstract
geometrical objects will have an explicit physical meaning.
So, one can think of $A_{\mu\nu}$ as the local angular velocity tensor of the
frame of reference. The first curvature vector of the congruence $F_\mu$
determines the acceleration of the reference body in a given point, and
$S_{\mu\nu}$ is the rate of strain tensor ~\cite{Ant}.

\section{\ab ~decomposition with respect to an adopted basis}

To find all the relations considered above in an $(n+m)$ decomposed form for some
fixed basis is a question of great significance for applications.
Coordinate form of $(n+m)$ decomposition considered in SecV is rather
cumbersome, and the objects themselves prove to be singular. One of the
reasons of this is that the range of indices
$\mu^{\prime},\mu^{\prime\prime},... $ is redundant. Therefore it is more
convenient for applications to choose adopted bases of $(n+m)$ decomposition
which will eliminate such redundancy. One's choice
of one basis or another is dictated by a physical situation, requirements of
an interpretation of results, or just by the necessity to use the most
comfortable way of calculation. We shall present here the invariant relations
of SecIII with respect to an adopted basis of decomposition. Note that
in such a form the formulae will be quite feasible for any concrete basis of
decomposition. All the known types of decomposition (for torsion free theories) can be obtained
as special cases of the present formalism by choosing the corresponding bases.
In an $(n+m)$ decomposed form our method is essentially useful for calculation of
the Riemann tensor, the Ricci tensor, and the curvature scalar by computer.

We shall now consider two adopted dual bases of decomposition: a vector one
$\{E_\mu\}=\{E_a ,E_i\}$ on $T(M)$, and a covector basis
$\{\theta^\mu\}=\{\theta^a ,\theta^i\}$ on $T^{*}(M)$,
where $E_b \in\Sigma^{\prime}\equiv \Sigma^n ,~
E_i \in\Sigma^{\prime\prime}\equiv \Sigma^m ;~
\theta^a \in\Sigma^{*\prime}\equiv \Sigma^{*n} ;~
\theta^i \in\Sigma^{*\prime\prime}\equiv \Sigma^{*m}$.
According to (\ref{29}-\ref{30}) we have
\begin{equation}\label{63}
\theta^a (E_b) = \delta^{a}_{b};~~~
\theta^a (E_j) =0;~~~
\theta^i (E_b) =0;~~~
\theta^i (E_k) =\delta^{i}_{k}
\end{equation}
\begin{equation}\label{64}
E_b\cdot E_k=0, \qquad <\theta^a,\theta^i> =0 \quad (a,b=1,2,...,n;\, i,k=n+1,n+2,...,n+m).
\end{equation}
It should be emphasized that the indices $a,b,c,...$ and $i,j,k,...$ indicate
the subbundles $\Sigma^n ,\Sigma^{*n}$ and $\Sigma^m ,\Sigma^{*m}$
respectively. With respect to the basis $\{E_\mu\},\{\theta^\mu\}$ one has
\begin{equation}\label{65}
    H^{\prime}=E_{a}\otimes \theta^{a};~~~
H^{\prime\prime}=E_{i}\otimes \theta^{i};~~~
g= g^{\prime}+g^{\prime\prime} =
\gamma_{ab}\theta^{a}\otimes\theta^{b} +
h_{ik}\theta^{i}\otimes\theta^{k}
\end{equation}
where $\gamma_{ab}=E_{a}\cdot E_{b}$ and $h_{ik}=E_{i}\cdot E_{k}$ are the
components of the metrics $g^{\prime},g^{\prime\prime}$ induced on the
subbundles $\Sigma^n$ and $\Sigma^m$.

Then we introduce the definitions
\begin{eqnarray}
      & \nabla^{\prime}_{E_a}E_b = L^{c}_{ab}E_c ;~~~~~~~
\nabla^{\prime\prime}_{E_i}E_j = L^{k}_{ij}E_k ; \nonumber \\
      & B^{\prime}(E_i ,E_k ) = B^{a}_{ik}E_a ;~~~~~
B^{\prime\prime}(E_a ,E_b ) = B^{i}_{ab}E_i
\label{66}
\end{eqnarray}
\begin{eqnarray}
      && [E_a ,E_b ]^{\prime} = \lambda^{c}_{ab}E_c ;~~~~~
[E_i ,E_j ]^{\prime\prime} = \lambda^{k}_{ij}E_k ; \nonumber  \\
      && [E_a ,E_i ]^{\prime} = \lambda^{b}_{ai}E_b ;~~~~~
[E_i ,E_a ]^{\prime\prime} = \lambda^{k}_{ia}E_k
\label{67}
\end{eqnarray}
where $L^{c}_{ab}$ and $L^{i}_{jl}$ are the coefficients of connections
$\nabla^{\prime}$ induced on $\Sigma^n$ and $\nabla^{\prime\prime}$ induced on
$\Sigma^m$. Similarly $B^{c}_{ik}$ and $B^{i}_{ab}$ are the coefficients of
the tensors of extrinsic non-holonomicity of the subbundles $\Sigma^m$ and
$\Sigma^n$ respectively. Using the identity (\ref{12}) one can find
\begin{eqnarray}
&&L^{c}_{ab} = \triangle^{c}_{ab} + \gamma^{c}_{ab};~~~~~
L^{i}_{jk} = \triangle^{i}_{jk} + \gamma^{i}_{jk};  \nonumber  \\
&&B^{a}_{ik} = S^{a}_{ik} + A^{a}_{ik};~~~~~
B^{i}_{ab} = S^{i}_{ab} + A^{i}_{ab}
\label{68}
\end{eqnarray}
where
\begin{equation}\label{69}
    2\triangle_{cab} =
E_a \gamma_{bc} + E_b \gamma_{ac} - E_c \gamma_{ab};~~~~~
2\gamma_{cab} =
\lambda_{cab} + \lambda_{bca} - \lambda_{abc}
\end{equation}
\begin{eqnarray}
     && 2S_{aik} = ({\cal L}_{E_a}g^{\prime\prime})(E_i ,E_k ) =
E_a h_{ik} + \lambda_{ika} + \lambda_{kia};  \nonumber  \\
     && 2A^{a}_{ik} = (d\theta^{a})(E_i ,E_k );~~~~
2A_{aik} = -E_{a}\cdot [E_i ,E_k ].
\label{70}
\end{eqnarray}

The coefficients $A_{iab},~S_{iab},~\gamma_{ijk},~\triangle_{ijk}$, unwritten
here, can be obtained from (\ref{69})-(\ref{70}) by the replacement
$(a,b,c,...\leftrightarrow i,j,k,...)$. Adhering to this style
here and below we shall write and discuss only those relations which can not
be found by the change of indices. We should remind also that the indices
$(a,b,c,...)$ are raised and lowered by the metrics $\gamma^{ab}$ and $\gamma_{ab}$.
 The curvature tensor and its contractions are presented in Appendix B.

In the special case of $(n+1)$ decomposition, i.e. when $m=1$ one has adopted
bases $\{E_\mu\}=\{E_a ,E\}~$, $\{\theta^\mu\}=\{\theta^a ,\theta\}$,
$~(a,b=1,2,...n)$, so that
\begin{eqnarray}
     && \theta^{a}(E_b ) = \delta_{a}^{~b};~~~~~
\theta_{a}(E) = 0 = \theta (E_a ); \nonumber  \\
     && \theta (E) = 1;~~~~
E\cdot E_a = 0;~~~~E\cdot E\equiv \varepsilon N^2
\label{71}
\end{eqnarray}
where $\{E_a\}\in \Sigma^n;~~\theta^{a}\in \Sigma^{*n}$ and
$E\in\Sigma^1 ;~\theta\in\Sigma^{*1}$.
In this case the projectors $H^{\prime}=E_{a}\otimes \theta^{a}$ and
$H^{\prime\prime}=E\otimes \theta$ induce the decomposition of the metric
\begin{equation}\label{72}
g= g^{\prime}+g^{\prime\prime} =
\gamma_{ab}\theta^{a}\otimes\theta^{b} +
\varepsilon N^2 \theta\otimes\theta.
\end{equation}
Then using the relations (\ref{66})-(\ref{70}), (\ref{B1})-(\ref{B9})
when $i=j=k=1$ or (\ref{48})-(\ref{54})
when $u=N^{-1}E,~~\omega =N\theta$ we can find all the necessary relations in
the $(n+1)$ decomposed form in an adopted basis. Thus, from (\ref{48}) it follows that
\begin{equation}\label{73}
F = N^{-2}(G - (E\log N)E);~~~~~~G = \nabla_{E}E.
\end{equation}
The tensor of extrinsic non-holonomicity of the subbundle $\Sigma^n$ can be
written in the form
\begin{eqnarray}\label{74}
B(E_a,E_b) = \varepsilon S_{ab}+A_{ab}
\equiv \varepsilon N^{-1}{\cal B}_{ab};~~~~~
2{\cal B}_{ab} = 2D_{ab} + F_{ab}
\end{eqnarray}
where
\begin{eqnarray}
  && S_{ab} = N^{-1}D_{ab};~~
2D_{ab} =({\cal L}_{E}g^{\prime})(E_a ,E_b ) =
E\gamma_{ab} + E_a\cdot [E_b ,E ] + E_b\cdot [E_a ,E ];  \nonumber \\
  && 2A_{ab} = \varepsilon N^{-1}F_{ab};~~~~
F_{ab} = \varepsilon N^2 (d\theta)(E_a ,E_b ) = -E\cdot [E_a ,E_b ].
\label{75}
\end{eqnarray}
Acting in the same way as in the previous sections we can find the
generalized Gauss-Codazzi-Ricci's equations (see Appendix C).

\section{Canonical parameterization of an \ab ~split structure \newline
          and its special cases}

The relations of SecVI are invariant under the transformation of adopted
bases:
\begin{eqnarray}
\theta^{a} = L^{a}_{b}e^{b};~~~~
\theta^{l} = L^{l}_{k}e^{k};~~~~
E_a = (L^{-1})^{~b}_{a}e_{b};~~~~
E_i = (L^{-1})^{~k}_{i}e_{k}.
\label{76}
\end{eqnarray}
where $\{L^{a}_{b}\}$ and $\{L^{k}_{i}\}$ are $(n\times n)$ and $(m\times m)$
non-singular matrices, and $\{(L^{-1})^{~b}_{a}\}$ and $\{(L^{-1})^{~k}_{i}\}$
are their inverse matrices. Using this property of invariance one can choose,
without loss of generality, the simplest basis of decomposition which
is useful for applications.

For this purpose we consider the expansion of the covector basis on
$\Sigma^{*m}$ in the domain ${\rm U}$ of definition of the map
$x^{\mu}~~(\mu = 1,2,...n,n+1,...n+m)$, i.e.
$\theta^{i} = \theta^{i}_{\mu}dx^{\mu}$. Due to the fact that the rank of the
$n\times (n+m)$ matrix $\{\theta^{i}_{\mu}\}$ is equal to $n$, there is an
$(m\times m)$ non-singular matrix $\{\theta^{i}_{k}\}$ as a box in
$\{\theta^{i}_{\mu}\}$. Then the covectors $\theta^{i}$
can be written in the form:
$\theta^{i} = \theta^{i}_{k}dx^k + \theta^{i}_{a}dx^a =
L^{i}_{k}(dx^k + N^{k}_{a}dx^a)\equiv L^{i}_{k}e^k$
where
$L^{i}_{k} = \theta^{i}_{k},~~N^{k}_{a} = (L^{-1})^{k}_{i}\theta^{i}_{a}$.
Thus the covector basis $\theta^{i}$ goes over into the new covector basis
$e^{k}\in\Sigma^{*m}$. The vector basis on $\Sigma^n$ can be written
similarly as $E_a = E^{~\mu}_{a}\partial_{\mu}$. From the condition of duality
$e^{k}(E_a ) = 0$ it follows that
$E_a =(L^{-1})^{b}_{a}(\partial_b -N^{k}_{b}\partial_k )\equiv
(L^{-1})^{b}_{a}e_b$, where
$(L^{-1})^{~b}_{a} = E^{~b}_{a}$. Thereby we defined the new vector basis
$e_{b}\in\Sigma^{n}$. The other vector and covector bases
($e^{i}\in\Sigma^{m}$ and $e^{a}\in\Sigma^{*n}$ respectively) are defined by
the condition of duality up to $(n\cdot m)$ functions $B^{a}_{i}$.
As a result one obtains the following parameterization of the basis of decomposition:
\begin{eqnarray}
  && e^{a} = dx^a + B^{a}_{i}e^i \in\Sigma^{*n};~~~~~~~
e_{a} = \partial_a - N^{i}_{a}\partial_i \in\Sigma^{n};  \nonumber \\
  && e^{i} = dx^i + N^{i}_{a}dx^a \in\Sigma^{*m};~~~~~
e_{i} = \partial_i - B^{a}_{i}e_{a} \in\Sigma^{m}.
\label{77}
\end{eqnarray}
We shall call this parameterization the canonical one.

If one follows similar procedure beginning with the covector basis
$\theta^{a}\in\Sigma^{*n}$, one will obtain the other canonical
parameterization of $(n+m)$ decomposition.
\begin{eqnarray}
      e^{a} = dx^a &+ A^{a}_{i}dx^i \in\Sigma^{*n};~~~~~
e_{a} = \partial_a - M^{k}_{a}e_{k} \in\Sigma^{n};  \nonumber \\
      e^{i} = dx^i &+ M^{i}_{a}e^{a} \in\Sigma^{*m};~~~~~
e_{k} = \partial_k - A^{a}_{k}\partial_a \in\Sigma^{m}.
\label{78}
\end{eqnarray}
When some metric $g$ is fixed on $M$, the functions $B^{a}_{i}$
(or $M^{i}_{a}$) can be found from the condition of orthogonality (\ref{64})
in terms of
$g_{\mu\nu}$ and $N^{i}_{a}$ (or $A^{b}_{k}$). If, otherwise, we fix
$B^{a}_{i}$ (or $M^{i}_{a}$), then we can obtain the metric for both cases
according to (\ref{65}):
\begin{eqnarray}
     && g = \gamma_{ab}(dx^a + B^{a}_{i}e^i)\otimes(dx^c + B^{c}_{k}e^k) +
h_{ik} e^i \otimes e^k   \nonumber \\
     && g = \gamma_{ab}e^a \otimes e^b + h_{ik}(e^i + M^{i}_{a}e^a)\otimes
(e^k + M^{k}_{b}e^b).
\label{79}
\end{eqnarray}
With respect to the canonically parameterized basis (\ref{77}),
the objects (\ref{68})-(\ref{70}) and the Lie bracket of the basic vector
fields have the form
\begin{eqnarray}
\label{80}
&& \lambda^{c}_{ab} = -2B^{c}_{i}A^{i}_{ab};~~~~~~
\lambda^{k}_{ij} = (B^{a}_{i}e_{j} - B^{a}_{j}e_{i})N^{k}_{a};  \nonumber \\
&& \lambda^{c}_{ai} = - e_a B^{c}_{i} + 2A^{k}_{ab}B^{b}_{i}B^{c}_{k} +
N^{k}_{a,i}B^{c}_{k};~~~~
\lambda^{k}_{ia} = -2A^{k}_{ac}B^{c}_{i} - N^{k}_{a,i}  \nonumber \\
&& 2A^{i}_{ab} = e_{b}N^{i}_{a} - e_{a}N^{i}_{b};~~~~
2A^{a}_{ij} = e_{i}B^{a}_{j} - e_{j}B^{a}_{i} -
\lambda^{k}_{ij}B^{a}_{k}                                  \\
&& 2S_{aik}  = ({\cal L}_{e_a}h)(e_{i},e_{k});~~~~~~
2S_{iab}  = ({\cal L}_{e_i }\gamma)(e_{a},e_{b}) \nonumber
\end{eqnarray}
where $\gamma =\gamma_{ab} e^a \otimes e^b$ and $h=h_{ik} e^i \otimes e^k$.
Here all the geometrical characteristics are expressed in terms of the
functions $h_{ij},\gamma_{ab},B^{a}_{i},N^{k}_{b}$ and their derivatives.
Substituting the
objects (\ref{80}) for those used in (\ref{B2})-(\ref{B8}) we can
obtain the Riemann tensor, the Ricci tensor and the scalar curvature in an
$(n+m)$ decomposed form with respect to the canonically parameterized basis
(\ref{77}). All the relations  for the parameterization (\ref{78}) are found
from (\ref{80}) by the substitution
$(a,b~\leftrightarrow ~i,j;~~B^{a}_{i}\to M^{i}_{a},~~N^{i}_{a}\to A^{a}_{i})$.

In the case of $(n+1)$ decomposition both types of parameterizations should be
considered independently. Thus for the $(3+1)$ monad method there are two
kinds of canonical parameterizations (with respect to local coordinates
$\{x^{\mu}\} = \{t,x^i \}$) determined by
\begin{eqnarray}
&& e_{0} = \partial_t - N^{i}\partial_{i} = Nu, ;~~~~
  e^{0} = dt + B_{i}e^i  = N^{-1}\omega ;  \nonumber \\
&& e_{i} = \partial_i - B_{i}e_0 ;~~~~~~~~~~~~~~~
  e^{i} = dx^i + N^{i}dt
\label{81}
\end{eqnarray}
and
\begin{eqnarray}
&&e_{0} = \partial_t - M^{i}e_{i} = Vu, ;~~~~
e^{0} = dt + A_{i}dx^i  = V^{-1}\omega ; \nonumber \\
&&e_{i} = \partial_i - A_{i}\partial_t ;~~~~~~~~~~~~~~~
e^{k} = dx^k + M^{k}e^0
\label{82}
\end{eqnarray}
where $u$ is a monad vector, $\omega$ is a one-form of time such that
$\omega(u)=1$.

The first set of bases (\ref{81}) is the generalization of the well-known ADM
parameterization ~\cite{ADM}.
In this case the metric has the form
\begin{equation}
ds^2 = N^2(dt + B_{j}e^j)^2 - h_{ik}e^i e^k ,~~~~(e^{i} = dx^i + N^{i}dt).
\label{83}
\end{equation}
The second set of bases (\ref{82}) implies that the metric is given by
\begin{equation}
ds^2 = V^2(e^0 )^2 - h_{ik}(dx^i + M^{i}e^0 )(dx^k + M^{k}e^0 ),
\label{84}
\end{equation}
where $e^{0} = dt + A_{j}dx^j$.

The latter parameterization is the generalization of those often used when
describing stationary spaces.
It is worth emphasizing that the redundant "degrees of freedom" of the
metrics (\ref{83})-(\ref{84}) may be used to fix a frame of reference or to simplify
the Einstein equations.
In the theory of stationary configurations, representation (\ref{84}) is useful for
examining of solutions, for which a flux of matter and the timelike Killing's
vectors are non-collinear (so-called skew solutions ~\cite{Kram}).

If $B_j$ vanishes the metric (\ref{83}) goes over into the standard
ADM parameterization
\begin{equation}
ds^2 = N^2 dt^2 - h_{ik}(dx^i + N^{i}dt)(dx^k + N^{k}dt).
\label{85}
\end{equation}
When $M^k$ vanishes, the metric (\ref{84}) has the form
\begin{equation}
ds^2 = V^2(dt + A_{j}dx^j )^2 - h_{ik}dx^i dx^k.
\label{86}
\end{equation}
This parameterization is often used when describing stationary spaces.
If we take $N^i =0$ or $A_j =0$ for the metrics (\ref{85}) and (\ref{86})
respectively, then in both cases we have
\begin{equation}
ds^2 = g_{00}dt^2 - h_{ik}dx^i dx^k.
\label{87}
\end{equation}
This kind of decomposition corresponds to a trivial case when $\Sigma^3$ is a
family of hypersurfaces, where each of hypersurfaces is orthogonal to the
curves $x^i ={\rm const}$. This decomposition is invariant under the transformations
\begin{equation}
t = t(t^{\prime}),~~x^i = x^i (x^{\prime k}).
\label{88}
\end{equation}
The three-dimensional part of these transformations acts uniformly on all the
hypersurfaces.
Now we shall start, otherwise, from three-dimensional transformations
(\ref{88}) which can be extended to the gauge ones by supposing that they
depend on time, i.e.
\begin{equation}
t = t(t^{\prime}),~~x^i = x^i (t^{\prime},x^{\prime k}).
\label{89}
\end{equation}
These transformations, under which the hypersurfaces $t={\rm const}$ remain
unchanged, have been called the kinemetric ones ~\cite{Zel2}. In order
that the decomposition of the metric be
invariant with respect to (\ref{89}) we must "make longer" the time derivative
$\partial_t \to\partial_t - N^{i}\partial_i $ (simultaneously we take
$dx^i \to dx^i + N^{i}dt$) by using the gauge vector $N^{i}$. Thus it leads
to the kinemetric method of decomposition ~\cite{Zel2}, which coincides with the
ADM representation ~\cite{ADM}.

Similarly extending the transformations of time we obtain the chronometric
transformations ~\cite{Zel1}
\begin{equation}
t = t(t^{\prime},x^{k\prime }),~~x^i = x^i (x^{\prime k}).
\label{90}
\end{equation}
It is obvious that the transformations (\ref{90}) do not change the congruence of
world lines $x^i = {\rm const}$. These transformations have been taken as a basis
for the definition of the frame of reference ~\cite{Zel1}. "Making longer" the time
differential
$dt \to dt + A_i dx^i$ (herewith $\partial_i \to\partial_i - A_{i}\partial_t$)
we obtain the chronometric method of decomposition. The transformations (\ref{90})
and (\ref{89}) are the complements of one another and form together the
general covariant transformations $x^{\mu} = x^{\mu}(x^{\prime\nu})$.

Further generalizations of (\ref{85}) and (\ref{86}) lead to various
parameterizations of the monad method. Thus, making longer
$dt,~dt \to dt + B_i e^i~(\partial_i \to\partial_i - B_{i}e_0 )$
one has the canonical parameterization (\ref{81}),(\ref{83}).
Making longer $\partial_t,~~\partial_t \to \partial_t - M^{i}e_i$
$~(dx^k \to dx^k + M^{k}e_0)$ one obtains the other canonical parameterization
(\ref{82}),(\ref{84}) of the monad
method for $M^4$. "Lengthening" as referred to is connected with
extension of the admissible transformations, which are not coordinate but
basic ones. The generalization (\ref{83}) of the ADM
parameterization is invariant under the transformations:
\begin{equation}
{\tilde e}^k = {\alpha}^{k}_{i}e^{i},~~~~
{\tilde h}_{ij} = {\alpha}^{m}_{i}\alpha^{n}_{j}h_{mn},~~~~
{\tilde B}_{j} = {\alpha}^{k}_{j}B_{k}.
\label{91}
\end{equation}
If we write the inverse of the metric (\ref{84})
\begin{equation}
(\partial_s )^2 = V^{-2}(\partial_t - M^{i}e_i )^2 - h^{ik}e_i e_k
\label{92}
\end{equation}
then it is easily can be seen that the metric (\ref{84}) is invariant under the
transformations
\begin{equation}
{\tilde e}_i = \beta^{k}_{i}e_{k},~~~~
{\tilde h}^{ij} = \beta^{i}_{m}\beta^{j}_{n}h^{mn},~~~~
{\tilde M}^{i} = \beta^{i}_{k}M^{k}.
\label{93}
\end{equation}
In (\ref{91}), (\ref{93}) $\{{\alpha}^{i}_{k}\}$ and $\{\beta^{i}_{k}\}$ are
non-singular matrices depending on a point $p$ of $M$.
From this we can clearly see the role of the parameterizations (\ref{83}),
(\ref{84}) as such generalizations of the kinemetric and chronometric methods that the
corresponding metrics admit non-holonomic transformations of spatial vector
and covector bases (\ref{91}) and (\ref{93}) respectively.

\section{Decomposition induced by a family of surfaces}

Let $\{M^m \subset M\}$ be an $n-$parameter family of $m-$dimensional surfaces.
One may think of these surfaces as intersections of the hypersurfaces
$x^a = {\rm const}$ i.e. $M^m =\bigcap_a \{x^a = {\rm const}\}$.
It is obvious that such a family induces $(n+m)$ decomposition of $M$. Indeed,
there exists the vector basis
$e_i = \partial_i$ on $T(M^* ),~(i=n+1,...,n+m)$, because of holonomicity
of the $M^m$ itself.
As a consequence of it, the covector basis on the orthogonal to $T(M^m )$
subbundles $\Sigma^n$ is a set of one-forms $\{e^a = dx^a\}$. The corresponding
dual bases to the bases $\{e_i\}$ and $\{e^a\}$ are determined up to $(n\cdot m)$
functions $N^{i}_{a}$ such that
\begin{eqnarray} \label{94}
&&e^a = dx^a \in\Sigma^{*n},~~~~~~
e_a = \partial_a - N^{i}_{a}\partial_i \in\Sigma^{n}  \nonumber  \\
&&e^i = dx^i + N^{i}_{a}dx^a \in\Sigma^{*m},~~~~~
e_i = \partial_i \in\Sigma^{m} .
\end{eqnarray}
The functions $N^{i}_{a}$ are expressed in terms of the components of the
metric $g$ by using the condition of orthogonality $e_{a}\cdot e_{i}=0$. Thus
the projection operators and the metric have the form:
\begin{equation}\label{95}
H^{\prime} = (\partial_a - N^{i}_{a}\partial_i )\otimes dx^a,~~~~~
H^{\prime\prime} = \partial_k \otimes(dx^k + N^{k}_{a}dx^a )
\end{equation}
\begin{equation}\label{96}
g = \gamma_{ab}dx^a \otimes dx^b +
h_{ik}(dx^i + N^{i}_{c}dx^c)\otimes (dx^k + N^{k}_{d}dx^d).
\end{equation}
From the form of the metric (\ref{96}) it can be seen that here we used the special
case of canonical parameterization
of $(n+m)$ decomposition (\ref{77}) when $B^{a}_{i}$ vanishes. In this case
the formulae (\ref{80}) become much simpler. Thus, one finds
\begin{equation}\label{98}
\lambda^{c}_{ab} = 0;~~~
\lambda^{k}_{ij} = 0;~~~
\lambda^{c}_{ai} = 0;~~~
\lambda^{k}_{ia} =  - N^{k}_{a,i};~~~
A^{a}_{ij} = 0;
\end{equation}
\begin{equation}\label{99}
2S_{iab}  = \gamma_{ab,i};~~~~~
2A^{i}_{ab} = e_{b}N^{i}_{a} - e_{a}N^{i}_{b}
\end{equation}
\begin{equation}\label{100}
2S_{aik}  = h_{ik,a} - h_{ik,l}N^{l}_{a} -
h_{lk}N^{l}_{a,i} - h_{il}N^{l}_{a,k}
\end{equation}
\begin{equation}\label{101}
2L_{cab} = 2\triangle_{cab} =
e_{a}\gamma_{bc} + e_{b}\gamma_{ca} + e_{c}\gamma_{ab}
\end{equation}
\begin{equation}\label{102}
2L_{ijk} = 2\triangle_{ijk} =
h_{ij,k} - h_{ik,j} - h_{jk,i}.
\end{equation}
The partial derivatives with respect to coordinates $x^i$ and $x^a$ are
denoted here by $",i"$ and $",a"$ respectively. Then, according to
(\ref{B2})-(\ref{B8}),
one can find the curvature tensor and its contractions.

\section{Decomposition induced by a group of isometries}

Let $M$ admits a non-transitive group of isometries $G^n$ with the $n$ linearly
independent Killing's vectors $\{\xi_a\}$, which satisfy the relations
\begin{equation}\label{103}
[\xi_a ,\xi_b ] = C^{d}_{ab}\xi_d~~~~~(a,b,d = 1,2,...n)
\end{equation}
where the $C^{d}_{ab}$ are the structure constants and obey the Jacobi
identity $C^{c}_{[ab}C^{f}_{d]c}=0$ and the condition
$C^{c}_{ab}+C^{c}_{ba}=0$. In addition, the metric $g$ satisfies the Killing's
equations:
\begin{equation}\label{104}
    ({\cal L}_{\xi_a }g)(X,Y) =
\xi_a (X\cdot Y) - [\xi_a ,X]\cdot Y - X\cdot [\xi_a ,Y] =0,~~~~~
\forall X,Y\in T(M).
\end{equation}

The group $G^n$ decomposes $M$ into a family of $m-$codimensional
surfaces $\{M^n\}\subset M$, on which $G^n$ is simply transitive
($\{M^n\}$ are invariant manifolds).
Thus, we can say that the group $G^n$ induces $(n+m)$ decomposition of $M$
into the $m-$parameter family of $n-$dimensional surfaces of transitivity.
Then the subbundle $\Sigma^n = \bigcup T(M^n)$
is a union of the tangent bundles of the family $\{M^n\}$, and
$\Sigma^m$ is a union of all the $m-$dimensional directions, which are
tangent to $M$ and orthogonal to $T(M^n)$.

Now we shall start in the same way as in the previous section. Thus one may
think of the surfaces $M^n$ as an intersection of the invariant hypersurfaces
$\{x^i ={\rm const}\}$, i.e. $M^m =\bigcap_i \{x^i ={\rm const}\},~~(i=n+1,...n+m)$.
Moreover, one has $dx^i(\xi_a ) = \xi_a x^i =0$. This is obvious that the
invariant differential one-forms $dx^i$ can be chosen as a covector basis on
the subbundles $\Sigma^{*m}$. Then there exists the vector basis
$\{\partial_a\}\in T(M^n)$, so that $dx^i(\partial_a ) =0$ and
$\xi_a = \xi^{b}_{(a)}\partial_b$.
Having extended these bases to the "complete ones":
$\{dx^i\}\to \{dx^\mu\} = \{dx^a ,dx^i\}\in T^{*}(M)$ and
$\{\partial_a\}\to\{\partial_\mu\} = \{\partial_a ,\partial_i\}\in T(M)$,
where $dx^\mu (\partial_\nu) = \delta^{\mu}_{\nu}$ and
$[\xi_a ,\partial_i ]=0$,
we can define one-forms $\omega^a$ such that
\begin{equation}\label{105}
      \omega^a (\xi_b) = \delta^{a}_{b};~~~~
\omega^a (\partial_i ) = 0;~~~~{\cal L}_{\partial_i }\omega^a =0
\end{equation}
\begin{equation}\label{106}
{\cal L}_{\xi_a }\omega^b = -C^{b}_{ad}\omega^d ;~~~~
2d\omega^a = C^{a}_{bd}\omega^b \wedge\omega^d ~.
\end{equation}
Let us now introduce an auxiliary definition. We shall say that a split
structure ${\cal H}^2$ is compatible with a group of isometries if the
conditions of invariance of ${\cal H}^2$ are satisfied, i.e. if
\begin{equation}\label{107}
{\cal L}_{\xi_a }H^{\prime} = 0,~~~~
{\cal L}_{\xi_a }H^{\prime\prime} =0,~~~~(a=1,2,...n).
\end{equation}
Using (\ref{65}) and (\ref{103}) one can easily verify that for the other vector and
covector bases $\{E_k\}\in\Sigma^m $ and $\{\theta^{a}\}\in\Sigma^{*n}$
we have, respectively,
\begin{equation}\label{108}
{\cal L}_{\xi_a }\theta^b = -C^{b}_{ad}\theta^d ;~~~~~~
{\cal L}_{\xi_a }E_k =0.
\end{equation}
To concretize the basis of decomposition we take
$\theta^{a} = \theta^{a}_{\mu}dx^{\mu}$ and $E_i = E^{\mu}_{i}\partial_{\mu}$.
Then the conditions of duality $\theta^{a}(\xi_b ) = \delta^{a}_{b},~~
\theta^{a}(E_i ) = 0,~~dx^{k}(E_i ) = \delta^{k}_{i}$ determine these
bases up to $(n\cdot m)$ functions $A^{a}_{i}$. As a result the basis of $(n+m)$
decomposition has the form:
\begin{eqnarray} \label{109}
 &&\xi_a \in\Sigma^n ;~~~~~e^a = \omega^a +
      A^{a}_{i}dx^i \in\Sigma^{*n} \nonumber \\
 &&e_i = \partial_i - A^{a}_{i}\xi_a \in\Sigma^m ;~~~~dx^k \in\Sigma^{*m},~~~~
[\xi_a ,e_i ] =0.
\end{eqnarray}
The projection operators and the metric can be written as
\begin{equation}\label{110}
H^{\prime} = \xi_a \otimes(\omega^a + A^{a}_{i}dx^i );~~~~~
H^{\prime\prime} = (\partial_i - A^{a}_{i}\xi_a)\otimes dx^i
\end{equation}
\begin{equation}\label{111}
g = g^{\prime} + g^{\prime\prime} =
\gamma_{ab}(\omega^a + A^{a}_{i}dx^i )\otimes (\omega^b + A^{b}_{j}dx^j )+
h_{kl}dx^k \otimes dx^l\, .
\end{equation}
From the Killing's equations one finds
\begin{equation}\label{112}
\xi_a \gamma_{bc} - C^{d}_{ab}\gamma_{dc} - C^{d}_{ac}\gamma_{bd} =0;~~~
\xi_a A^{b}_{i} - C^{b}_{ad}A^{d}_{i} =0;~~~
\xi_a h_{ik} =0.
\end{equation}
Using these equations we obtain the main geometrical objects
\begin{eqnarray} \label{113}
&&A^{\prime\prime}(\xi_a ,\xi_b ) =0;~~~~
2A^{\prime}(e_i ,e_k )\equiv F^{a}_{ik}\xi_a \nonumber \\
&&F^{a}_{ik} = A^{a}_{k,i} - A^{a}_{i,k}
      + C^{a}_{bd}A^{b}_{k}A^{d}_{i} \nonumber \\
&&S^{\prime}(e_i ,e_k )=0;~~~~~
2e_i \cdot S^{\prime\prime}(e_a ,e_b )\equiv
2S_{iab}=e_i \gamma_{ab} \\
&&2L_{abc} = C_{cab}+C_{bca}+C_{acb}; \nonumber \\
&&2L_{ijk}=2\triangle_{ijk}=e_j h_{ik}+e_k h_{ij}-e_i h_{jk}. \nonumber
\end{eqnarray}
In the end, from (\ref{B2})-(\ref{B8}), we can find the curvature tensor, the
Ricci tensor and scalar curvature (see Appendix D). When $m=0$ we come to the
case of homogeneous spaces.

\section{Lagrangians of the unified multidimensional  \newline
            Kaluza-Klein theories}

The mathematical model we shall use for spaces of the unified theories is the
totality of the following objects:
      a) a connected $(4+n)-$dimensional pseudo-Riemannian $C^\infty$ manifold
$M^{4+n}$ with a non-singular metric $g$ on it;
      b) an $n-$parameter compact group of isometries $G^n$ on $M^{4+n}$ with
linearly independent Killing's vectors $\xi_a \in T(M^{4+n})$ for which the
structure constants $C^{a}_{bd}$ satisfy the condition
$C^{a}_{ad}=0,~(a,b,d=4,5,...n+3)$.

The physical space-time $V^4 \equiv M^{4+n}/M^n$ is the quotient space $M^{4+n}$
with respect to the invariant manifolds $M^n$ of the group $G^n$. The $V^4$ is
described by the components $h_{ik}$ of the metric $h$, by the set of gauge
fields $A^{b}_{i}$ and by the multiplet $n(n+1)/2$ of scalar fields
$\varphi_{ab}\equiv -\gamma_{ab}$. All these tensors are obtained under the
$(4+n)$ decomposition of $M^{4+n}$ (see SecIX). The true physical
configuration is described not by a single set of fields
$\{h_{ik},A^{b}_{j},\varphi_{cd}\}$, but by a whole equivalence class of such
sets; each of them corresponds to some point of the orbit $G^n$.
The signature of the metric $g$ is defined by two conditions: first,
the metric $h$ is a Lorentz
one, and second, the energy density is positive for obtained Lagrangian of
fields $\{A^{b}_{j},\varphi_{cd}\}$.
In addition, the metric $g$ satisfies the $(4+n)-$dimensional
variational Hilbert Principle for the functional $S[g]$, i.e.
\begin{equation}\label{114}
\delta S[g] = \delta\left\{-\frac{1}{4\pi V}\int R^{(4+n)}\Omega^{(4+n)}\right\} =0
\end{equation}
where $R^{(4+n)}$ is the curvature scalar on $M^{(4+n)}$, the $(4+n)-$form
$\Omega^{(4+n)}$ is the volume measure on $M^{4+n}$, and $V$ is the
$n-$dimensional invariant volume of $M^n$
\begin{equation}\label{115}
V = \int_{M^n}\omega^4 \wedge\omega^5\wedge...\wedge\omega^{4+n} \equiv
\int_{M^n}\Omega^{(n)}.
\end{equation}

The conditions $C^{a}_{ab}=0$ follow from the requirement that the
volume measure $\Omega^{(n)}$ must be invariant. They are necessary for compatibility of the
variational Hilbert Principle and homogeneity of $M^n$ with respect to the
group of isometries $G^n$. This restricts the admissible variations of fields
${\cal L}_{\xi_a }\delta g =0$ in (\ref{114}). (The similar situation may be
found in the theory of homogeneous models of cosmology ~\cite{Ryan},~\cite{Bog}).

Using the formulae of SecIX and Appendix D for the metric $g$ in the
$(n+4)$ decomposed form
\begin{equation}\label{116}
g = h_{ik}dx^{i}\otimes dx^{k} -
\varphi_{ab}(\omega^a + A^{a}_{m}dx^{m})\otimes (\omega^b + A^{b}_{n}dx^{n})
\end{equation}
and omitting a divergence of some vector, we obtain
\begin{equation}\label{117}
S^{(4+n)}[g] = S[\varphi_{ab},A^{a}_{i},h_{jk}] = \int_{V^4}\sqrt{-h}Ld^{4}x.
\end{equation}
The Lagrangian density is
\begin{eqnarray} \label{118}
  &&\sqrt{-h}L = - \frac{1}{4\pi}\sqrt{\mid h\varphi\mid}\{R^{(4)}
   + \frac{1}{4}\varphi_{ab}F^{a}_{ij}F^{bij} \nonumber \\
&&~~~~~~~~~~~~~~~~~~~~+ ~(\varphi^{ab}\varphi^{cd} - \varphi^{ac}\varphi^{bd})
h^{ik}D_i \varphi_{ab}D_k \varphi_{cd} + U(\varphi_{ab})\}
\end{eqnarray}
where
\begin{equation}\label{119}
U(\varphi_{ab}) =\frac{1}{2} \varphi^{cd}C^{a}_{bc}(C^{b}_{ad} +
\frac{1}{2}\varphi_{ap}\varphi^{bq}C^{p}_{qd})
\end{equation}
and
\begin{equation}\label{120}
D_i \varphi_{ab} = \varphi_{ab,i} - T(A_i )^{d}_{a}\varphi_{db} -
T(A_i )^{d}_{b}\varphi_{ad}
\end{equation}
is the gauge-invariant derivative.
The components $T(A)^{a}_{b}\equiv C^{a}_{bd}A^d$ of the matrix $T(A)$ realize
the adjoint representation of the group $G^n$:
$[T(A),T(B)] = T([A,B]),~A = A^a\xi_a ,~B = B^a\xi_a $.
Lagrangian of this kind (but with the second derivatives of the fields
$\varphi_{ab}$) has been obtain in ~\cite{Cho}.

When $n=1$ Lagrangian (\ref{118}) reduces to Lagrangian of the
5-dimensional Kaluza-Klein Theory ~\cite{Gl1}. In the
static case of spherical symmetry from $n=1$ it follows
Lagrangian of the simple dynamic system. Its equations can be integrated
by separation of variables of the corresponding Hamilton-Jacobi equation.
In such a way the solution for the interacting scalar, electromagnetic, and
gravitational fields was obtained in ~\cite{Gl5}
within the framework of the Unified 5-dimensional Kaluza-Klein Theory.

\section{Relativistic configurations of a perfect fluid}

Let us consider space-time $M^4$ with the metric $g$ in the $(3+1)$ decomposed
form
\begin{equation}\label{121}
g = V^2 e^{0}\otimes e^{0} - h_{ik}e^{i}\otimes e^{k},~~~~
g^{-1} = V^{-2}e_{0}\otimes e_{0} - h^{ik}e_{i}\otimes e_{k}
\end{equation}
where $g^{-1}$ is the inverse of the metric $g$. For the time being, we require
the basis of decomposition to be an adopted abstract one (i.e. not concretized). Let the
source of the gravitational field described by the metric (\ref{121}) be a
perfect fluid with the field of 4-velocities
$u = V^{-1}e_{0} = d/ds$ which is tangent to the flow lines
$x^{\mu}=x^{\mu}(s)$. Herewith the mass density $\rho$ obeys the conservation law:
\begin{equation}\label{122}
{\rm div} (\rho u)\equiv (\nabla_{e_\mu}\rho u)(e_\mu) =
V^{-1}h^{-1/2}{\cal L}^{\prime\prime}_{e_0 }(\rho h^{1/2}) =0
\end{equation}
where ${\cal L}^{\prime\prime}_{e_0 }$ is the Lie
derivative with respect to the basis $~\{e_i\}$:
${\cal L}^{\prime\prime}_{e_0 }\sqrt{h}
=\frac{1}{2}\sqrt{h}h^{ik}({\cal L}_{e_0 }h)(e_i ,e_k )$.
The equation of motion for the fluid follows from the relation:
\begin{equation}\label{123}
{\rm div} T \equiv (\nabla_{e_\mu}T)(e^\mu ,~.~)=0.
\end{equation}
The energy-momentum tensor $T$ is
\begin{equation}\label{124}
T = \mu V^{-2}e_{0}\otimes e_{0} + Ph^{ik}e_{i}\otimes e_{k}
\end{equation}
where $\mu$ is the energy density of the fluid, $P$ is the pressure. Using the
thermodynamic relations
\begin{equation}\label{125}
d{\cal H} = Tds + \rho^{-1}dP,~~~~~~{\cal H} = (\mu +P)\rho^{-1}
\end{equation}
one finds the equations of motion
\begin{equation}\label{126}
({\rm div} T)(e_0 ) = \rho TV^{-1}uS = -\rho V^{-1}dS/ds =0
\end{equation}
\begin{equation}\label{127}
({\rm div} T)(e_i ) = h^{ik}(dP - \rho{\cal H}{\cal L}_{u}\omega)(e_{k}) =0.
\end{equation}
Here we use the following notations: ${\cal H}$ is the enthalpy, $S$ is the
entropy, $T$ is the temperature, and $\omega$ is the covector of the 4-velocity
of the fluid  ($\omega = Ve^{0},~\omega(u)=1$).
We introduce "the one-form of the enthalpy $\theta$" and "the two-form of the
curl $\Omega$" by
\begin{equation}\label{128}
\theta = {\cal H}\omega = {\cal H}Ve^0 ,~~~~~~\Omega = d\theta .
\end{equation}
Then the equations of motion (\ref{126}), (\ref{127}) can be expressed as
\begin{equation}\label{129}
{\cal L}_{e_0 }\theta = d({\cal H}V) -VTdS.
\end{equation}
Using the formula ${\cal L}_{e_0 } = i_{e_0 }d + di_{e_0 }$, where the operator
$i_{e_0 }$ is defined by the relation
$(i_{e_0 }\Omega )(Y) = \Omega (e_0 ,Y),~\forall Y\in T(M^4 )$, we obtain one
more form of the equations of motion
\begin{equation}\label{130}
i_{e_0 }\Omega = -VTdS.
\end{equation}
The condition of integrability of these relations leads to the equations of
motion for the curl of a perfect fluid
\begin{equation}\label{131}
{\cal L}_{e_0 }\Omega = -d(TV)\wedge dS.
\end{equation}
In the special case $S={\rm const}$ a perfect fluid is isentropic so that
the equations for "the one-form of the enthalpy" (\ref{129}) and
"the two-form of the curl" (\ref{130}), (\ref{131}) are reduced to
the relations:
\begin{equation}\label{132}
{\cal L}_{e_0 }\theta = d({\cal H}V)
\end{equation}
\begin{equation}\label{133}
i_{e_0 }\Omega =0,~~~~~~{\cal L}_{e_0 }\Omega = 0.
\end{equation}
It is to be note that the last equation in (\ref{133}) is the condition of
integrability of the equation (\ref{132}). Moreover we may regard this
condition as an invariant formulation of the theorem ~\cite{Sibg}, which
states that the
two-form of the curl $\Omega$ is constant along the world lines of particles
of an isentropic perfect fluid. From the first relation in (\ref{133}) it
follows that $\Omega$ is singular, i.e.
$\Omega(e_0 ,X)=0,~\forall X\in T(M^4 )$,
and therefore "completely spatial". This implies
\begin{equation}\label{134}
\Omega = \sum_{i,j}\Omega_{ij}e^{i}\wedge e^{j};~~~~~~
\Omega\wedge\Omega = d\theta\wedge d\theta =0.
\end{equation}
Since in general case $\theta\wedge d\theta \not= 0$, then according to
the theorem Darboux (see, for example ~\cite{Ster}) it
follows that there exist such functions $\xi ,\eta ,\zeta$ that
$\theta = d\xi +\eta d\zeta$. This representation has been used in
~\cite{Kras} to construct a number of families of solutions of the
Einstein equations for an isentropic perfect fluid.

Now we shall consider the stationary spaces of General Relativity with a
timelike Killing's vector $\partial_t$. Then the equations (\ref{126}),
(\ref{127}), as well as their consequences (\ref{129})-(\ref{133}), go over
into the equilibrium conditions of a perfect fluid. For an isentropic
stationary flow they admit completely $3-$dimensional formulation.
Indeed, in this case one has
\begin{equation}\label{135}
{\cal L}_{\partial_t }g =0,~~~~{\cal L}_{\partial_t }e^\mu =0,~~~~
[\partial_t ,e_\mu] =0.
\end{equation}

Then using the parameterization of decomposition (\ref{82}) we deduce that the functions
$V,A_i ,M^k , h_{ik}$ as well as $\rho ,\mu ,P,{\cal H}$ do not depend
on time. We define the vector $\vec{M}$ and covector $A$ on the subbundles
$\Sigma^{\prime\prime}\equiv\Sigma^{3}$ by
\begin{equation}\label{136}
\vec{M} = M^i\partial_i ,~~~~~~A = A_k dx^k .
\end{equation}
In terms of $\vec{M}$ and $A$ the conservation law for mass (\ref{122})
is transformed into the $3-$dimensional equation of continuity of the flow
lines
\begin{equation}\label{137}
{\rm div} ^{(3)}(\rho\vec{M})\equiv (\nabla_{e_i}\rho\vec{M})(e^i )=
h^{-1/2}{\cal L}_{\vec{M}}(\rho h^{1/2}) =0.
\end{equation}
When $S={\rm const}$ the condition (\ref{129}) may be rewritten in the
$3-$dimensional form as well
\begin{equation}\label{138}
i_{\vec{M}}dA = -d\log{({\cal H}V)};~~~~\vec{M}({\cal H}V) =0.
\end{equation}
From now on the objects and operations are defined on the $3-$dimensional
manifold $t={\rm const}$ with respect to the bases $\{\partial_i\}$ and $\{dx^k\}$.
For example: $dA = (1/2){\cal F}_{ik}dx^i \wedge dx^k$, where
${\cal F}_{ik}=A_{k,i}-A_{i,k}$. The equilibrium condition (\ref{138}) may be
expressed in the form
\begin{equation}\label{139}
{\cal L}_{\vec{M}}A = d\{ A(\vec{M}) - \log{({\cal H}V)} \}
\end{equation}
showing that the one-form ${\cal L}_{\vec{M}}\vec{A}$ is exact. Hence, as the
condition of integrability one obtains the conservation
$3-$dimensional theorem for the curl $dA$ along the $3-$dimensional flow lines, i.e.
\begin{equation}\label{140}
{\cal L}_{\vec{M}}dA =0.
\end{equation}

In the case of parameterization (\ref{81}) for the stationary spaces the
functions $V,B_i ,N^k ,h_{jk}$ do not depend on time either. By analogy with
(\ref{139}) one has
\begin{equation}\label{141}
{\cal L}_{\vec{N}}B = -d\log{({\cal H}V)}
\end{equation}
where
\begin{equation}\label{142}
\vec{N}=N^i\partial_i ,~~~~B = B_k dx^k .
\end{equation}

The condition of integrability gives the conservation theorem for the
curl of $B$
\begin{equation}\label{143}
{\cal L}_{\vec{N}}dB =0.
\end{equation}

If one of the two objects $A$ and $\vec{M}$ in (\ref{139}) (or $\vec{N}$ and
$B$ in (\ref{141})) vanishes then the equilibrium condition of an isentropic
perfect fluid has the simple form
\begin{equation}\label{144}
{\cal H}V = V(\mu +p)/\rho =k
\end{equation}
where $k$ is the constant. Thus the Lagrangian of an isentropic perfect fluid
in equilibrium is
\begin{equation}\label{145}
L_m \equiv -V\sqrt{h}P = (k\rho - \mu V)\sqrt{h} =
[k - (1+\varepsilon)V]\rho\sqrt{h}
\end{equation}
where $\varepsilon =\varepsilon(\rho)$ is the internal energy of the fluid
and $\mu =\rho(1+\varepsilon)$.

As was noted above, the parameterizations (\ref{81}), (\ref{82}) have spurious degrees of
freedom. It means that the vector $\vec M$ or covector $A$ in (\ref{82}) can be chosen
arbitrarily, by using additional physical reasons. Therefore we have a right to
introduce the potential of rotation $\Psi_1$ by the formula
\begin{equation}\label{146}
{\cal L}_{\vec{M}}A = d(\log\Psi_1 ).
\end{equation}
Then the equilibrium condition (\ref{139}) is written as a relation for potentials
\begin{equation}\label{147}
\Psi_1 HV = C_1 {\rm exp}(A_i M^i),~~~~C_1 = {\rm const}     \nonumber
\end{equation}
and actually gives us the integral of motion.
In another case of the parameterization the equilibrium condition
(\ref{81}) can be expressed in the form
\begin{equation}\label{148}
{\cal L}_{\vec{N}}B = d(\log\Psi_2 ),~~~~~\Psi_2 HV = C_2 = {\rm const}.
\end{equation}
Thus the potentials $\Psi_1$ and $\Psi_2$ are different from each other by
the exponential factor ${\rm exp}(A_i M^i)$.


\begin{acknowledgments}
We would like to acknowledge M.Korkina and  A.Sokolovsky for interesting
and helpful discussions, and Yu.Vladimirov for
encouragement and continual attention to the theme of this research.
The authors wish also to thank M.Godina  who was helpful in pointing out
some inaccuracies and A.Boroviec for drawing our attention to several
important works on the theory of "almost-product structures".
\end{acknowledgments}

\appendix
\section{The generalized Gauss-Codazzi-Ricci's equations}

Replacing all the connections in the definition of the curvature
tensor (\ref{15}) by their "split representatives" (\ref{16})-(\ref{19}) we have obtained
the invariant non-holonomic generalizations of the Gauss-Codazzi-Ricci's
equations:
\begin{eqnarray}\label{A1}
    R(X^{a},Y^{a})Z^{a}\cdot V^{a} &=&
R^{a}(X^{a},Y^{a})Z^{a}\cdot V^{a} +
\sum_{c\not=a}\{2A^{c}(X^{a},Y^{a})\cdot
B^{c}(Z^{a},V^{a})                   \nonumber \\
&+& ~B^{c}(Y^{a},V^{a})\cdot
B^{c}(X^{a},Z^{a})-
B^{c}(X^{a},V^{a})\cdot
B^{c}(Y^{a},Z^{a})\},
\end{eqnarray}
\begin{eqnarray}\label{A2}
    R(X^{a},Y^{a})Z^{a}\cdot V^{b} &=&
V^{b}\cdot \{(\nabla^{b}_{Y^{a}} B^{b})(X^{a},Z^{a})-
(\nabla^{b}_{X^{a}} B^{b})(Y^{a},Z^{a})\}    \nonumber \\
&+& ~2Z^{a}\cdot B^{a}(A^{b}(X^{a},Y^{a}),V^{b})+
\sum_{c\not=a,b}\{2Z^{a}\cdot Q^{a}
(A^{c}(X^{a},Y^{a}),V^{b})                 \nonumber      \\
&+& ~B^{c}(X^{a},Z^{a})\cdot Q^{c}(Y^{a},V^{b})-
B^{c}(Y^{a},Z^{a})\cdot Q^{c}(X^{a},V^{b})\}
\end{eqnarray}
\begin{eqnarray}\label{A3}
    R(X^{a},Y^{b})Z^{a}\cdot V^{b} &=&
(Z^{a}\cdot (\nabla^{a}_{X^{a}} B^{a})+
<X^{a}\cdot B^{a},Z^{a}\cdot B^{a}>)(Y^{b},V^{b})   \nonumber\\
&+& ~(V^{b}\cdot (\nabla^{b}_{Y^{b}} B^{b})+
<Y^{b}\cdot B^{b},V^{b}\cdot B^{b}>)
(X^{a},Z^{a})                                       \nonumber\\
&+& ~\sum_{c\not=a,b}\{B^{c}(X^{a},Z^{a})\cdot
B^{c}(Y^{b},V^{b})-
Q^{c}(X^{a},V^{b})\cdot
Q^{c}(Y^{b},Z^{a})                                  \nonumber\\
&+& ~V^{b}\cdot Q^{b}
(\Lambda^{c}(X^{a},Y^{b}),Z^{a})\}
\end{eqnarray}
\begin{eqnarray}\label{A4}
    R(X^{a},Y^{b})Z^{a}\cdot V^{d} &=&
V^{d}\cdot\{(\nabla^{d}_{Y^{b}}B^{d})(X^{a},Z^{a})-
(\nabla^{d}_{X^{a}} Q^{d})(Y^{b},Z^{a})        \nonumber\\
&-& ~B^{d}(Y^{b},B^{b}(X^{a},Z^{a}))\}+
Z^{a}\cdot\{B^{a}(Y^{b},Q^{b}(X^{a},V^{d}))     \nonumber\\
&-& ~B^{a}(\Lambda^{d}(X^{a},Y^{b}),V^{d})\}
+ (<Y^{b}\cdot B^{b},V^{d}\cdot B^{d}>)(X^{a},Z^{a}) \nonumber\\
&-& ~(<X^{a}\cdot B^{a},V^{d}\cdot Q^{d}>)(Y^{b},Z^{a})-
\sum_{c\not=a,b,d}\{Z^{a}\cdot Q^{a}
(\Lambda^{c}(X^{a},Y^{b}),V^{d})                 \nonumber \\
&-& ~B^{c}(X^{a},Z^{a})\cdot Q^{c}(Y^{b},V^{d})+
Q^{c}(X^{a},V^{d})\cdot Q^{c}(Y^{b},Z^{a})\}
\end{eqnarray}
\begin{eqnarray}\label{A5}
    R(X^{a},Y^{b})Z^{c }\cdot V^{d} &=&
V^{d}\cdot\{(\nabla^{d}_{Y^{b}} Q^{d})(X^{a},Z^{c})-
(\nabla^{d}_{X^{a}} Q^{d})(Y^{b},Z^{c})\}        \nonumber \\
&+& ~B^{d}(X^{a},Q^{a}(Y^{b},Z^{c}))-
B^{d}(Y^{b},Q^{b}(X^{a},Z^{c}))+
B^{d}(\Lambda^{c}(X^{a},Y^{b}),Z^{c})\}          \nonumber \\
&+& ~Z^{c}\cdot\{B^{c}(Y^{b},Q^{b}(X^{a},V^{d}))-
B^{c}(X^{a},Q^{a}(Y^{b},V^{d}))                  \nonumber \\
&-& ~Q^{c}(\Lambda^{c}(X^{a},Y^{b}),V^{d})\}+
(<Y^{b}\cdot B^{b},V^{d}\cdot Q^{d}>)(X^{a},Z^{c}) \nonumber \\
&-& ~(<X^{a}\cdot B^{a},V^{d}\cdot Q^{d}>)(Y^{b},Z^{c})+
\sum_{f\not=a,b,c,d}
\{Q^{f}(Y^{b},V^{d})\cdot Q^{f}(X^{a},Z^{c})     \nonumber \\
&-& ~Q^{f}(Y^{b},Z^{c})\cdot Q^{f}(X^{a},V^{d})+
V^{d}\cdot Q^{d}(\Lambda^{f}(X^{a},Y^{b}),Z^{c})\}.
\end{eqnarray}
In the formula (\ref{A1}) the curvature tensor $R^a$ of the subbundle
$\Sigma^a$, introduced in \cite{Gl4}, is
\begin{eqnarray}\label{A6}
    R^{a}(X^{a},Y^{a})Z^a \equiv
\{\nabla^{a}_{X^a}\nabla^{a}_{Y^a} -
\nabla^{a}_{Y^a}\nabla^{a}_{X^a} - \nabla^{a}_{[X^{a},Y^{a}]^a} +
2\sum_{c\not=a}{\cal L}^{a}_{A^{c}(X^{a},Y^{a})}\}Z^a.
\end{eqnarray}
The covariant derivatives of the values $B^d$ and $Q^d$ are given by
\begin{eqnarray}\label{A7}
   (\nabla^{b}_{X^{a}} B^{b})(Y^{a}, Z^{a}) =
 \nabla^{b}_{X^{a}} (B^{b} (Y^{a}, Z^{a})) -
 B^{b}(\nabla^{a}_{X^{a}} Y^{a}, Z^{a}) -
 B^{b}(Y^{a},\nabla^{a}_{X^{a}} Z^{a})
\end{eqnarray}
\begin{eqnarray}\label{A8}
   (\nabla^{d}_{X^{b}} B^{b})(Y^{a}, Z^{a}) =
 \nabla^{d}_{X^{b}} (B^{d} (Y^{a}, Z^{a})) -
 B^{d}(\nabla^{a}_{X^{b}} Y^{a}, Z^{a}) -
 B^{d}(Y^{a},\nabla^{a}_{X^{b}} Z^{a})
\end{eqnarray}
\begin{eqnarray}\label{A9}
   (\nabla^{d}_{X^{a}} Q^{d})(Y^{b}, Z^{c}) =
 \nabla^{d}_{X^{a}} (Q^{d} (Y^{b}, Z^{c})) -
 Q^{d}(\nabla^{b}_{X^{a}} Y^{b}, Z^{c}) -
 Q^{d}(Y^{b},\nabla^{c}_{ X^{a}} Z^{c}).
\end{eqnarray}
We also used the definition
\begin{eqnarray}\label{A10}
   (<Y^{b}\cdot B^{b},V^{d}\cdot Q^{d}>)(X^{a}, Z^{c})\equiv
   <Y^{b}\cdot B^{b}(X^{a},~.~),V^{d}\cdot Q^{d}(~.~,Z^{c})> .
\end{eqnarray}
When fixing the vectors $X^{a},~Y^{b},~V^{d},~Z^{c}$, the definition
(\ref{A10}) gives us the scalar product $<\alpha^{ba},\beta^{dc}>$ of
one-forms
$$
      \alpha^{ba}\equiv Y^{b}\cdot B^{b}(X^{a},~.~),~~~~
      \beta^{dc} \equiv V^{d}\cdot Q^{d}(~.~,Z^{c})
$$

When $n_a =1~~(a=1,2,...r)$, i.e. when all the subbundles are one-dimensional,
the relations obtained here reduce to the $r$-dimensional variant of the
tetradic method's formulae ~\cite{Land}.

\section{Components of the curvature tensor with respect to an adopted
         basis for \ab ~decomposition}

Due to the definitions
\begin{eqnarray}\label{B1}
   \{E_\mu\}=\{E_a ,E_i\};~~
R(E_\mu ,E_\nu )E_\rho\cdot E_\sigma = R_{\sigma\rho\mu\nu};~~
R(E_\mu ,E_\nu )E_\rho = R^{\sigma}_{\rho\mu\nu}E_\sigma
\end{eqnarray}
the generalized Gauss-Codazzi-Ricci's equations (\ref{39})-(\ref{42})
have the form
\begin{eqnarray}\label{B2}
R_{abcd} = R^{(n)}_{abcd} + 2A^{i}_{.cd}B_{iba} +
B^{i}_{.cb}B_{ida} + B^{i}_{.db}B_{ica}
\end{eqnarray}
\begin{eqnarray}\label{B3}
   R_{ibcd} = B_{icb\mid d} - B_{idb\mid c} + 2A^{k}_{.cd}B_{bki} +
B^{k}_{.db}(B_{cik} - \lambda_{kic}) -
B^{k}_{.cb}(B_{dik} - \lambda_{kid})
\end{eqnarray}
\begin{eqnarray}\label{B4}
    R_{ibcj} = B_{bji\mid c} - B_{icb\mid j}
- B_{bjk}B^{~~k}_{ci.} - B_{icd}B^{~~d}_{jb.}
 + B_{bki}\lambda^{k}_{.jc} + B_{bjk}\lambda^{k}_{.ic} +
B_{idb}\lambda^{d}_{.cj} + B_{icd}\lambda^{d}_{.bj}
\end{eqnarray}
where the curvature tensor of the subbundle $\Sigma^n$ is defined by its
components $R^{(n)}_{abcd}$ according to
\begin{eqnarray}\label{B5}
    R^{(n)a}_{~~~~bcd} = E_c L^{a}_{db} - E_d L^{a}_{cb} +
L^{f}_{db}L^{a}_{cf} - L^{f}_{cb}L^{a}_{df} -
\lambda^{f}_{cd}L^{a}_{fb} + 2A^{i}_{.cd}\lambda^{a}_{bi}
\end{eqnarray}
(and similarly for the replacement  $n\rightarrow m$ and
$a,b,c,...\leftrightarrow i,j,k,...$).
Then the components of the Ricci tensor and the curvature scalar have the form
\begin{eqnarray}\label{B6}
    R_{bd} = R^{(n)}_{bd} - B^{i}_{db\mid i} &-& S_{b\mid d} +
2A^{i}_{ad}A^{~~a}_{ib.} + 2S_{iad}S^{ia}_{~~b} \nonumber \\
&-& ~S^{~ij}_{b}S_{dij} + A_{bij}A^{~ij}_{d} - S_{i}B^{i}_{.db} -
B^{i}_{.da}\lambda^{a}_{.bi} - B^{i}_{.ab}\lambda^{a}_{.di}
\end{eqnarray}
\begin{eqnarray}\label{B7}
    R_{ia} = B^{~~~b}_{ia.~\mid b} + B^{~~~k}_{ai.~\mid k}
&-& S_{i\mid a} - S_{a\mid i} -
2S^{b}_{ik}S^{k}_{ab} - 6A^{b}_{ik}A^{k}_{ab}        \nonumber \\
&+& ~S^{k}(B_{aik} - \lambda_{kia}) + S^{b}(B_{iab} - \lambda_{bai}) +
B^{k}_{ab}\lambda^{~~b}_{ki.} + B^{b}_{ik}\lambda^{~~k}_{ba.}
\end{eqnarray}
\begin{eqnarray}\label{B8}
    R = R^{(n)} - 2S^{i}_{~\mid i} - S^{i}S_{i} &-&
S^{i}_{ab}S^{ab}_{i..} - A^{i}_{ab}A^{ab}_{i..}      \nonumber\\
&+& ~R^{(m)} - 2S^{a}_{~\mid a} - S^{a}S_{a} -
S^{a}_{ij}S^{ij}_{a.} - A^{a}_{ij}A^{~~ij}_{a..}
\end{eqnarray}
where $S^{i}=S^{i}_{ab}\gamma^{ab},~~S^{a}=S^{a}_{ik}h^{ik}$.
The signs "$_{\mid i}$" and "$_{\mid a}$" denote the covariant derivative with
respect to the connections $L^{k}_{mn}$ and $L^{a}_{bc}$ in the directions of
the vectors $E_i$ and $E_a$ respectively. For example
\begin{eqnarray}\label{B9}
B_{icb\mid d} = E_d B_{icb} - B_{iab}L^{a}_{dc} - B_{ica}L^{a}_{db}~~~~
(a,b,c \leftrightarrow i,j,k).
\end{eqnarray}
The other components of the Ricci tensor and the curvature tensor can be
found from (\ref{B2})-(\ref{B7}) by the formal substitution
$a,b,c,...$ for $i,j,k,...$ and otherwise.

\section{Components of the curvature tensor with respect to an adopted
         basis for \cd ~decomposition}

The generalized Gauss-Codazzi-Ricci's equations for the metric (\ref{72})
with respect to the basis (\ref{71}) have the form:
\begin{eqnarray}\label{C1}
    R_{abcd} = R^{(n)}_{abcd} +
\varepsilon N^{-2}({\cal B}_{cb}{\cal B}_{da} -
{\cal B}_{db}{\cal B}_{ca} + F_{cd}{\cal B}_{ba})
\end{eqnarray}
\begin{eqnarray}\label{C2}
    R_{n+1,bcd} = N\{(N^{-1}{\cal B}_{cb})_{\mid d} -
      (N^{-1}{\cal B}_{db})_{\mid c}\} - \varepsilon N^{-2}G_b F_{cd}
\end{eqnarray}
\begin{eqnarray}\label{C3}
    R_{n+1,bc,n+1} = N{\cal L}_{E}(N^{-1}{\cal B}_{cb}) -
      {\cal B}_{ca}{\cal B}_{b.}^{~a} +
\varepsilon N^{-2}G_b G_c - N^2(N^{-2}G_b)_{\mid c}
\end{eqnarray}
\begin{eqnarray}\label{C4}
    R_{bd} =
R^{(n)}_{bd} - \varepsilon N^{-2}[N{\cal L}_{E}(N^{-1}{\cal B}_{db}) +
      D{\cal B}_{db} &+& \frac{1}{2}F_{ba}F_{d.}^{~a} -
      2D_{ba}D_{d.}^{~a}] \nonumber \\
      &+& \varepsilon (N^{-2}G_b)_{\mid d} - N^{-4} G_b G_d
\end{eqnarray}
\begin{eqnarray}\label{C5}
    R_{n+1,a} =
N[(N^{-1}{\cal B}_{a.}^{~b})_{\mid b} - E_a (N^{-1}D)] -
\varepsilon N^{-1}F_{ab}G^b
\end{eqnarray}
\begin{eqnarray}\label{C6}
    R_{n+1,n+1} =
-NE(N^{-1}D) - D_{ab}D^{ab} + \frac{1}{4}F_{ab}F^{ab}
+~N^2(N^{-2}G^a )_{\mid a} - \varepsilon N^{-2}G_a G^a
\end{eqnarray}
\begin{eqnarray}\label{C7}
    R = R^{(n)} - 2\varepsilon N^{-1}E(N^{-1}D)
- \varepsilon N^{-2}(D^2 &+& D_{ab}D^{ab}+\frac{1}{4}F_{ab}F^{ab}) \nonumber\\
&+& 2\varepsilon (N^{-2}G_a)_{\mid a} - 2N^{-4} G_a G^a.
\end{eqnarray}
\begin{eqnarray}\label{C8}
    R^{(n)a}_{~~~~bcd} = E_c L^{a}_{db} - E_d L^{a}_{cb} +
L^{f}_{db}L^{a}_{cf} - L^{f}_{cb}L^{a}_{df} -
\lambda^{f}_{cd}L^{a}_{fb} + \varepsilon N^{-2}F_{cd}\lambda^{a}_{b}
\end{eqnarray}
were $\lambda^{a}_{b}=\theta^{a}([E_b,E])$ and
$R^{(n)}=\gamma^{bd}R^{(n)}_{~bd};~~R^{(n)}_{~bd} = R^{(n)a}_{~~~bad}$.

\section{Components of the curvature tensor for a decomposition induced by
         a group of isometries}

The curvature tensor and its contractions with respect to the basis (\ref{109})
for the metric (\ref{111}) have the form:
\begin{equation}\label{D1}
      R^{(m+n)}_{dcab} = R^{(n)}_{dcab} + S_{ic[a}S^{i}_{b]d}
\end{equation}
\begin{equation}\label{D2}
      R^{(m+n)}_{icab} = S^{k}_{c[a}F_{b]ki} + S_{icd}C^{d}_{.ba}
      + 2S_{id[a}\gamma^{d}_{.b]c}
\end{equation}
\begin{equation}\label{D3}
      R^{(m+n)}_{ickb} = -S_{ibc;k} + S_{ibd}S^{~~d}_{kc.}
      + \frac{1}{4}F_{ckj}F^{~~j}_{bi.}- \frac{1}{2}\gamma^{d}_{.bc}F_{dki}
\end{equation}
\begin{equation}\label{D4}
      R^{(m+n)}_{ajkl} = F_{aj[l;k]} + F_{bj[k}S^{~~b}_{l]a}
      + F_{bkl}S^{~b}_{ja}
\end{equation}
\begin{equation}\label{D5}
      R^{(m+n)}_{ijkl} = R^{(m)}_{ijkl} + \frac{1}{2}F_{ai[k}F^{a}_{.l]j} -
      \frac{1}{2}F_{aij}F^{a}_{kl}
\end{equation}
\begin{equation}\label{D6}
      R^{(m)i}_{jlk} = 2e_{[k}\triangle^{i}_{l]j}
      + 2\triangle^{m}_{j[l}\triangle^{i}_{k]m},~~~~
      R^{(n)d}_{~.cab} = 2\gamma^{~d}_{q.[a}\gamma^{~~~q}_{b]c.}
      - C^{qd}_{..c}\gamma_{aqb}
\end{equation}
\begin{equation}\label{D7}
      R^{(m+n)}_{ab} = R^{(n)}_{ab} - S^{i}_{ab;i} - S^{i}_{ab}S_{i} +
      2S^{i}_{~ac}S^{~c}_{i~b} + \frac{1}{4}F_{aij}F^{~ij}_{b}
\end{equation}
\begin{equation}\label{D8}
      R^{(m+n)}_{ai} = \frac{1}{2}F^{~~k}_{ai~;k} + \frac{1}{2}F_{ail}S^{l}
      + C^{d}_{.ba}S^{b}_{id} - C^{b}_{bd}S^{~d}_{ia}
\end{equation}
\begin{equation}\label{D9}
      R^{(m+n)}_{ik} = R^{(m)}_{ik} - S_{(i;k)} - S_{iab}S^{~ab}_{k} +
      \frac{1}{2}F_{aij}F^{aj}_{..k}
\end{equation}
\begin{equation}\label{D10}
      R^{(m+n)} = R^{(n)} + R^{(m)} - 2S^{i}_{~;i} - S^{i}S_{i} -
      S^{iab}S_{iab} - \frac{1}{4}F^{a}_{ij}F^{~ij}_{a}.
\end{equation}
Here $R^{(m)} = h^{ik}R^{(m)}_{~ik};~~R^{(m)}_{~ik} = R^{(m)l}_{~~~ilk}$ and
$R^{(n)}=\gamma^{bd}R^{(n)}_{~bd};~~R^{(n)}_{~bd} = R^{(n)a}_{~~~bad}$.
The covariant derivative in the direction of the vector
$e_{k}$ with respect to the connection $\triangle^{i}_{jk}$ is denoted by
$";k"$.

\def\CMPh{Commun. Math. Phys.}
\def\JPh{J. Phys.}
\def\CJP{Czech. J. Phys.}
\def\LMPh {Lett. Math. Phys.}
\def\NPh  {Nucl. Phys.}
\def\PhE  {Phys.Essays}
\def\PhL  {{\it Phys. Lett.}~}
\def\PhR  {{\it Phys. Rev.}~}
\def\PhRL {Phys. Rev. Lett.}
\def\PhRp {Phys. Rep.}
\def\NCim {Nuovo Cimento}
\def\NuPB {Nucl. Phys.}
\def\GRG {{\it Gen. Relativ. Gravit.}~}
\def\CQG {Class. Quantum Grav.}
\def\prp {report}
\def\Prp {Report}
\def\GrC {{\it Gravitation$\&$Cosmology}~}
\def\DANS {{\it Dokl.Akad.Nauk SSSR}~}
\def\APh {{\it Ann.Phys.}~}
\def\JMM {{\it Journ.Math. and Mech.}~}
\def\JMP {{\it J.Math. Phys.}~}
\def\IVUZ {{\it Izv.Vyssh.Uchebn.Zaved.Fiz}~}
\def\APP {{\it Acta Phys.Pol.}~}

\def\jn#1#2#3#4#5{{#1}{#2} {\bf #3}, {#4} {(#5)}}

\def\boo#1#2#3#4#5{{\it #1} ({#2}, {#3}, {#4}){#5}}

\def\prpr#1#2#3#4#5{{``#1,''} {#2 }{#3}{#4}, {#5} (unpublished)}

\end{document}